\documentclass[12pt]{article}
\usepackage{amsfonts,color}
\usepackage{graphicx}
\usepackage{epsfig}
\usepackage{amsmath}
\usepackage{amssymb}
\usepackage{multirow}
\usepackage{hyperref}
\usepackage{slashed,cancel}
\begin{document}

\thispagestyle{empty}

%\begin{center}
%\hfill SINP%\\
%\end{center}

\begin{center}
\vspace{1.7cm}
{\Large\bf{ 
RS resonance in di-final state production at the LHC to NLO+PS accuracy}}
\vspace{1.4cm}

Goutam Das~$^{a,}$\footnote{goutam.das@saha.ac.in}\ ,
%\\[.5cm]
Prakash Mathews~$^{a,}$\footnote{prakash.mathews@saha.ac.in}\ ,
%\\[.5cm]
V. Ravindran~$^{b,}$\footnote{ravindra@imsc.res.in}\ ,
%\\[.5cm]
Satyajit Seth~$^{a,}$\footnote{satyajit.seth@saha.ac.in}\
\\[.5cm]
${}^a$ Saha Institute of Nuclear Physics, \\
 1/AF Bidhan Nagar, Kolkata 700 064, India
\\[.5cm]
${}^b$The Institute of Mathematical Sciences, \\
Tharamani, Chennai 600 113, India
\\[1cm]
\end{center}
\vfill

\begin{abstract}

We study the di-final state processes ($\ell^+ \ell^-$, $\gamma \gamma$,
$ZZ$, $W^+ W^-$) to NLO+PS accuracy, as a result of both the SM and RS
Kaluza-Klein graviton excitations.  Decay of the electroweak gauge
boson final states to different leptonic states are included at the
showering stage.  A selection of the results has been presented with PDF
and scale uncertainties for various distributions.  Using the di-lepton
and di-photon final states, we present the search sensitivity, for the
$14$ TeV LHC at $50$ fb$^{-1}$ luminosity.

\end{abstract}

\vfill
\newpage

\section{Introduction:}

With the discovery of a new scalar particle at the Large Hadron Collider
(LHC) and with the additional data on completion of Run I, it now appears
that the scalar particle is more and more likely to be the Standard Model
(SM) Higgs boson.  At the LHC, there has been no evidence of new physics
so far.  Nonetheless we know that the SM on many counts is not a
complete description of nature.  The SM can account for only a meager
4\% of the energy composition of the universe, does not account for
the observed phenomena in the neutrino sector.  Also the theoretical
criterion of naturalness needs the presence of physics beyond the SM
at the TeV scale.  As the properties of the Higgs boson become known
with greater precision, signs of new physics might show up.  We are
hence gearing up to the Run II with higher center of mass energies
at the LHC.  

Search of physics beyond the SM is an important objective of the
LHC physics program and is motivated by the large hierarchy that
exists between the gravitational Planck scale and the electroweak
symmetry breaking scale.  Among various options, an interesting
alternative that addressed the hierarchy problem was achieved by
invoking extra spacial dimensions in TeV scale brane world scenarios 
\cite{ArkaniHamed:1998rs,Antoniadis:1998ig,Randall:1999ee}.
Classification based on the geometry of extra spacial dimension
leads to two classes of model {\it viz.}\ the factorisable 
\cite{ArkaniHamed:1998rs,Antoniadis:1998ig} and non-factorisable extra
dimensions \cite{Randall:1999ee}.  In both these models the SM particles
are constrained in the 4-dimensional world and only gravity is allowed
in the extra dimensional bulk.  The ADD model with factorisable extra 
dimensions, has negligible curvature results in a tower of spin-2
Kaluza-Klein (KK) modes while the RS model with non-factorisable extra 
dimension has significant curvature leading to narrow spin-2 KK mode
resonances.  The phenomenology of these two models are quite
distinct, with the ADD model leading to an enhancement of the tale of
the invariant mass distribution as a result of the combined effect
of the tower of KK modes.  In contrast, the RS model leads to the
production of a narrow width RS KK mode resonances.

The geometry of the RS model consists of an extra spacial dimension $\phi$
which is warped, wherein two 3-branes are placed at orbifold fixed
points at $\phi=0$ (Planck brane) and $\phi=\pi$ (TeV brane).  The SM
particles are constrained on the TeV brane, while gravity originates on
the Planck brane and are allowed to propagate in the bulk.  The tower of
KK excitations ($h_{\mu\nu}^{(n)}$) of the graviton couples to the SM
energy momentum tensor ($T^{\mu\nu}$) through the following interaction
Lagrangian \cite{Davoudiasl:1999jd}, 
\begin{eqnarray}
{\cal L}_{int} &=& - \frac{1}{\overline{M_p}} T^{\mu\nu}(x)
h_{\mu\nu}^{(0)}(x) - 
\frac{e^{\pi{\cal{K}}R_c}}{\overline{M_p}} \sum_{n=1}^\infty
T^{\mu\nu}(x) h_{\mu\nu}^{(n)}(x) \qquad ,
\label{lagrangeRS}
\end{eqnarray}
where $\overline{M_p}$ is the reduced Planck scale, ${\cal{K}}$ is the
curvature assumed to be of the same order as $M_p$ and $R_c$ is the
radius of compactification.  The first term in the above Lagrangian
denotes the contribution of the zero mode graviton which is $M_p$
suppressed. However, due to the exponential warping the higher
dimensional Planck scale could be of the order of the electroweak scale.
As a consequence, it is customary to neglect the zero mode of graviton
excitation and consider the following interaction Lagrangian without
any loss of generality, 
\begin{eqnarray}
{\cal L}_{int}^{\mbox{\tiny{RS}}} &=& - \frac{\overline{c_0}}{m_0}
\sum_{n=1}^\infty T^{\mu\nu}(x) h_{\mu\nu}^{(n)}(x) \qquad ,
\end{eqnarray}
where $\overline{c_0} = {\cal{K}}/{\overline{M_p}}$ is an effective
coupling and $m_0 = {\cal{K}}e^{-\pi{\cal{K}}R_c}$ sets a mass scale
for the massive KK mode gravitons.  The 5-dimensional metric is
non-factorisable and this leads to a spectrum of KK modes whose masses
are given by, 
\begin{eqnarray}
 M_n = x_n~{\cal{K}}~e^{-\pi{\cal{K}}R_c} \qquad ,
 \label{gmass}
\end{eqnarray}
where $x_n$s' are the roots of the Bessel function $J_1(x)$.  The RS
model is characterised by the dimensionless coupling $\overline{c_0}$
and the first KK mode excitation mass $M_1$.

The KK modes could be produced {\it via} the $q \bar q$ channel or
the $gg$ channel which would then decay to SM bosons or fermions
leading to di-final states.  These processes are being explored at
the LHC leading to bounds on the model parameters
\cite{Chatrchyan:2011wq, Chatrchyan:2011fq, ATLAS:2011ab, Aad:2012cy}.
Of course, to put stringent bounds on the model parameters at a hadron
collider like the LHC, it is essential to have the next-to-leading order
(NLO) QCD corrections, as the leading order (LO) predictions suffer
from large theoretical uncertainties.  Presently the di-final
state processes are available to NLO accuracy for DY
\cite{Mathews:2004xp,Kumar:2006id}, di-photon \cite{Kumar:2009nn,Kumar:2008pk},
$ZZ$ \cite{Agarwal:2009xr,Agarwal:2009zg}, $W^+ W^-$
\cite{Agarwal:2010sp,Agarwal:2010sn} and di-jet \cite{Li:2014awa} for
the extra dimension scenarios ADD and RS.  This has been further extended
to the NLO+PS accuracy for the ADD model for the di-final state
processes \cite{Frederix:2012dp,Frederix:2013lga}, excluding the jets.
As the K-factor at NLO level is large, attempts to go beyond NLO
in QCD are underway and there are already first results 
\cite{deFlorian:2013sza,deFlorian:2013wpa,Ahmed:2014gla}
at NNLO level in the threshold limit for Drell-Yan production in
gravity mediated models.

In this paper, we present the di-final state production processes  
(except those containing jet(s) in the LO) at hadron colliders which
are interfaced with Parton Shower (PS) Monte Carlo to NLO accuracy
using the MC@NLO formalism for the RS model.
It should be pointed out that, the success to fully automatise the SM
calculations to this accuracy is rather recent \cite{Alwall:2014hca}, but the
status of BSM models to the same accuracy is still wanting.
Precise theoretical predictions to NLO+PS accuracy are extremely
desirable for the RS model and hence these codes are being made
available
for the LHC community.  Various physical observables are studied,
which are of relevance to future studies of these processes.
These processes have been probed at the LHC Run I, in the SM, Higgs
production and BSM searches and a more detailed study is expected
in the Run II.  

Rest of the paper is organised as follows: In section 2 we describe
the NLO results for the di-final state processes and match them to
parton shower Monte Carlo.  The numerical results are presented in
section 3 and finally we summarise our results in the last section.

\section{NLO with PS}

We have considered NLO QCD corrections to all the jet-exclusive tree level
di-final state processes, namely Drell-Yan, di-photon, $ZZ$ and $W^+W^-$ 
production processes in the SM as well as in the RS scenario and these
${\cal{O}}(\alpha_s)$ corrected results are then matched with parton shower 
Monte Carlo using MC@NLO formalism \cite{Frixione:2002ik}. The total RS
contribution represents both the signal and background together consisting
of pure SM, pure BSM and the interference between the two, whereas the SM
contribution alone is treated as background. 

For all the above mentioned processes, the parton level Born squared
amplitude in the SM comes only from $q\bar{q}$ initiated Feynman
diagrams, while in the RS model both $q\bar{q}$ and $gg$ initiated
Feynman diagrams contribute.  In addition there is interference
terms between the SM and RS model subprocesses.  In the fixed order analysis,
the ${\cal{O}}(\alpha_s)$ correction terms correspond to two categories
of Feynman diagrams: ({\it i}) real emission and ({\it ii}) one-loop
virtual correction. In the real emission part, $q\bar{q}$ or $gg$
initiated subprocesses contribute leading to an extra gluon emission 
in addition to the desired final state.  The $\bar{q}(q)g$ initiated
partonic subprocesses begin to contribute at ${\cal O} (\alpha_s)$.
Matrix elements coming from the
one-loop virtual diagrams participate in the ${\cal{O}}(\alpha_s)$
correction, when multiplied by the corresponding Born amplitude.
All these partonic subprocesses producing ${\cal{O}}(\alpha_s)$ correction
terms behave alike for all of our processes of interest. Moreover, one
additional ${\cal{O}}(\alpha_s)$ contribution shows up in the fixed order
calculation for the di-boson final state processes due to the interference
between $gg$ initiated box diagrams in the SM and $gg$ initiated Born
diagrams in the RS scenario. We have taken care of all the aforesaid
contributions in our present calculation. 

While dealing with Drell-Yan process, we have only considered $e^+e^-$
final state, as the other possible channel {\it i.e.}\, 
$PP\rightarrow \mu^+\mu^- X$ would be phenomenologically same with the
chosen one, apart from the experimental identification of the final state
particles.

In case of di-photon production, we have adopted smooth cone isolation
technique \cite{Frixione:1998jh} proposed by Frixione to get rid of
using fragmentation contribution which are non-perturbative in nature
which indicate the probability of fragmenting a parton into photon.  We call
it as Frixione isolation (FI) which ensures that soft radiation is not
eliminated in any region of phase space and at the same time guarantees
infra-red (IR) safety of the observable. In order to implement it, a cone
of radius 
$r=\sqrt{(\eta-\eta_{\gamma})^2 + (\phi-\phi_{\gamma})^2}$ is to be defined
centering around the direction of each photon in the pseudo-rapidity
($\eta$) and azimuthal angle ($\phi$) plane. It is then demanded that in order
to satisfy the isolation criteria, the sum of the hadronic transverse energy
$H(r)$ has to be always less than $H(r)_{\mbox{\scriptsize{max}}}$ for all
cones with radius $r\leq r_0$. In the present analysis, we have taken following
choice of $H(r)_{\mbox{\scriptsize{max}}}$ defined as, 
\begin{eqnarray}
 H(r)_{\mbox{\scriptsize{max}}} = \epsilon_{\gamma}\ E_T^{\gamma} 
\left(\frac{1-\cos{r}}{1-\cos{r_0}}\right)^n \quad , 
\end{eqnarray}
where $E_T^{\gamma}$ is the transverse energy of the photon and 
$\epsilon_{\gamma}$, $r_0$ and $n$ are three FI parameters that
are to be specified while applying such isolation.

On-shell $Z$ and $W^{\pm}$ have been produced while generating
events for $ZZ$ and $W^+W^-$ production processes respectively.
The two $Z$ bosons are then leptonically decayed to $e^+e^-$ and
$\mu^+\mu^-$ respectively at the time of showering.  The
decay channels $W^+\rightarrow e^+\nu_e$ and 
$W^-\rightarrow \mu^-{\bar{\nu}}_{\mu}$ have been taken into
 account while showering $W^+W^-$ events. 

Owing to the tremendous development in computation of NLO correction
in the last few years, automation plays an important role throughout
this work.  The universal FeynRules \cite{Alloul:2013bka} output
(UFO) of the RS model has been imported within the M{\sc ad}G{\sc raph}5
(MG5) environment \cite{Alwall:2011uj}.  
We choose to work in the MG5\_{\sc a}MC@NLO framework
\cite{Alwall:2014hca} in which the Born level square matrix elements
are generated using MG5 and calculation of the real emission
cross sections together with their singularities are overseen by
M{\sc ad}FKS \cite{Frederix:2009yq} package which uses the FKS
subtraction scheme \cite{Frixione:1995ms} in an automated way.
However, a set of external FORTRAN codes, that handle virtual
contributions, have been prepared using the analytical results
involving one-loop amplitudes for $e^+e^-$ \cite{Mathews:2005bw},
$\gamma\gamma$ \cite{Kumar:2009nn}, $ZZ$ \cite{Agarwal:2009zg},
$W^+W^-$ \cite{Agarwal:2010sn} in the SM \& RS model and they
have been systematically implemented into this framework.
Nevertheless, another in-house FORTRAN code which takes care of
the summation of KK-tower propagators, has suitably been fitted
in this environment, thereby making essential and appropriate
changes in the spin-2 HELAS routines \cite{Frederix:2012dp}. We
have explicitly checked numerical cancellation of double and
single poles coming from the real and virtual parts for all
these processes and thereafter used this complete set-up to
generate corresponding events.  The generated events are then
matched with HERWIG6 \cite{Corcella:2000bw} parton shower using
{\sc a}MC@NLO, where the MC@NLO formalism is being automatised.
Uncertainties in renormalisation ($\mu_R$) and factorisation
($\mu_F$) scale and in parton distribution functions (PDF) are
also estimated in an automated way with no extra CPU cost.  Note
that, instead of decaying a particle at the time of showering,
it is also possible to decay it into its preferred decay channel
at the event generation level itself by making the use of
M{\sc ad}S{\sc pin} \cite{Artoisenet:2012st} that restrains
nearly all spin correlations. However, to use the same for $Z$
or $W^{\pm}$ decay is beyond the scope of this paper due to the
significant complexity involved in including KK-tower summation and
changing HELAS \cite{Murayama:1992gi, deAquino:2011ix} routines
accordingly. 

\section{Numerical Results}

In this section, we present a number of differential distributions of
various kinematical observables at the NLO+PS accuracy for $e^+e^-$,
$\gamma\gamma$, $ZZ$, $W^+W^-$ production processes at the LHC with
center of mass energy $\sqrt{S}=14$ TeV. Following electroweak input 
parameters have been used at the time of event generation: ({\it i})
$\alpha_{EW}^{-1}=132.507$, ({\it ii}) $G_F=1.16639 \times 10^{-5}$
GeV$^{-2}$, ({\it iii}) $M_Z=91.188$ GeV and using them as input
the mass of $W$ boson $M_W=80.419$ GeV and $\sin^2{\theta_w}=0.222$
are evaluated.  We have considered $n_f=5$ massless quark flavors in
our present study. The central choices of $\mu_R$ and $\mu_F$ are
always set equal to the invariant mass of the corresponding di-final
state. MSTW(n)lo2008cl68 PDF sets \cite{Martin:2009iq} have been 
used throughout the analysis in order to generate (N)LO events and
they determine the value of strong coupling $\alpha_s$. (N)LO events
are generated with very loose cuts on transverse momentum ($P_T$)
and rapidity ($y$) of the final state particles and they are detailed
here under: ({\em a}) Drell-Yan: $P_T^{e^+,e^-} \geq 15$ GeV, 
$|y^{e^+,e^-}|\leq2.7$, $\Delta R^{e^+e^-} > 0.3$, where 
$\Delta R = \sqrt{(\Delta y)^2+(\Delta\phi)^2}$ denotes the separation
between two particles in the rapidity-azimuthal angle plane, ({\em b})
Di-photon: $P_T^{\gamma} \geq 15$ GeV, $|y^{\gamma}|\leq2.7$ with a set
of particular FI parameters {\it i.e.}, $\epsilon_\gamma=1$, $r_0=0.3$
and $n=2$. However, no such kinematical cuts have been provided while
generating events for $ZZ$ and $W^+W^-$ processes. Besides, for $W^+W^-$
event generation we have taken diagonal unit CKM matrix and neglected
top quark contribution in the whole analysis. 
\begin{figure}
 \centering{
 \includegraphics[height=8.0cm,width=7.4cm]{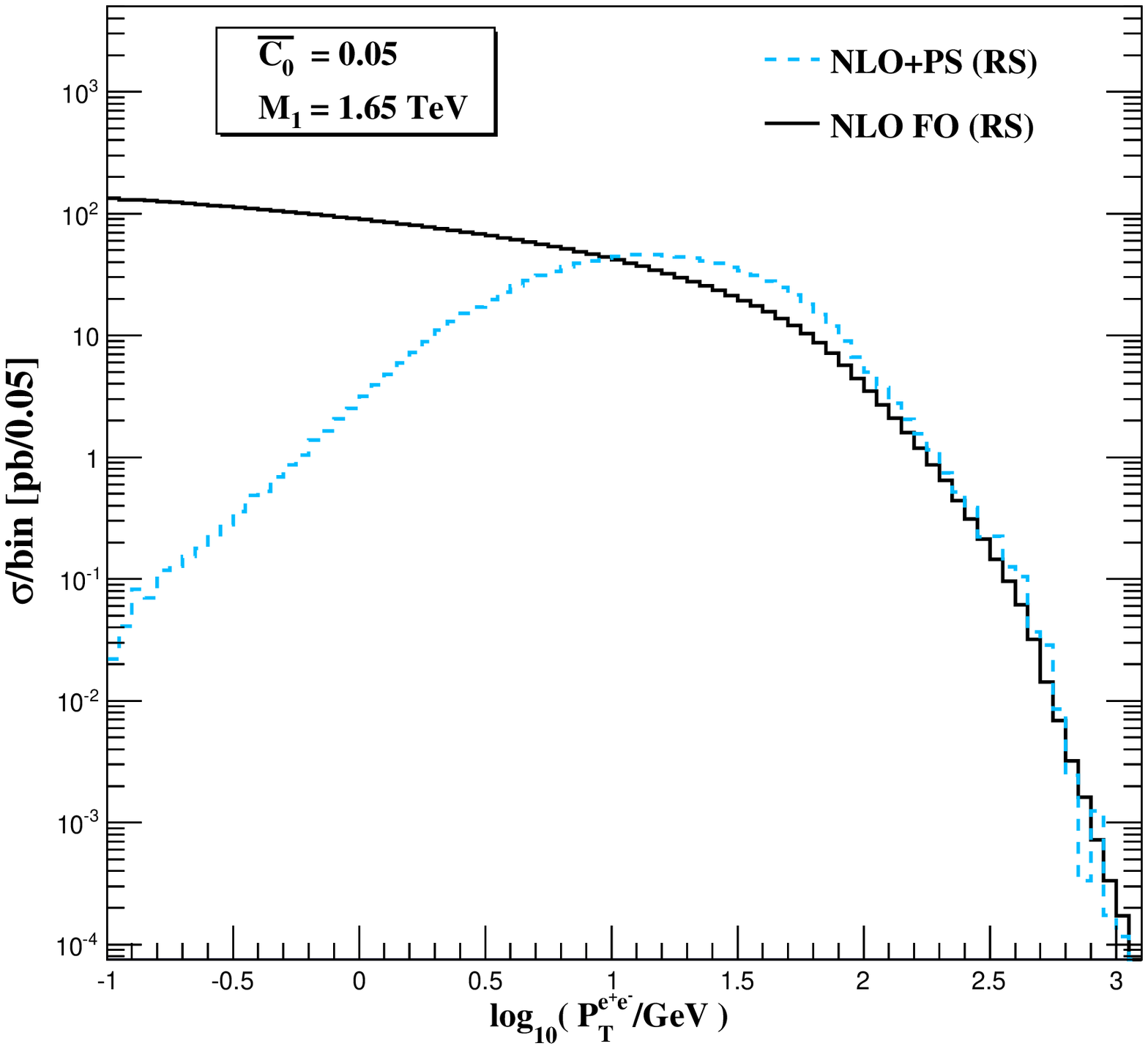}
 \includegraphics[height=8.0cm,width=7.4cm]{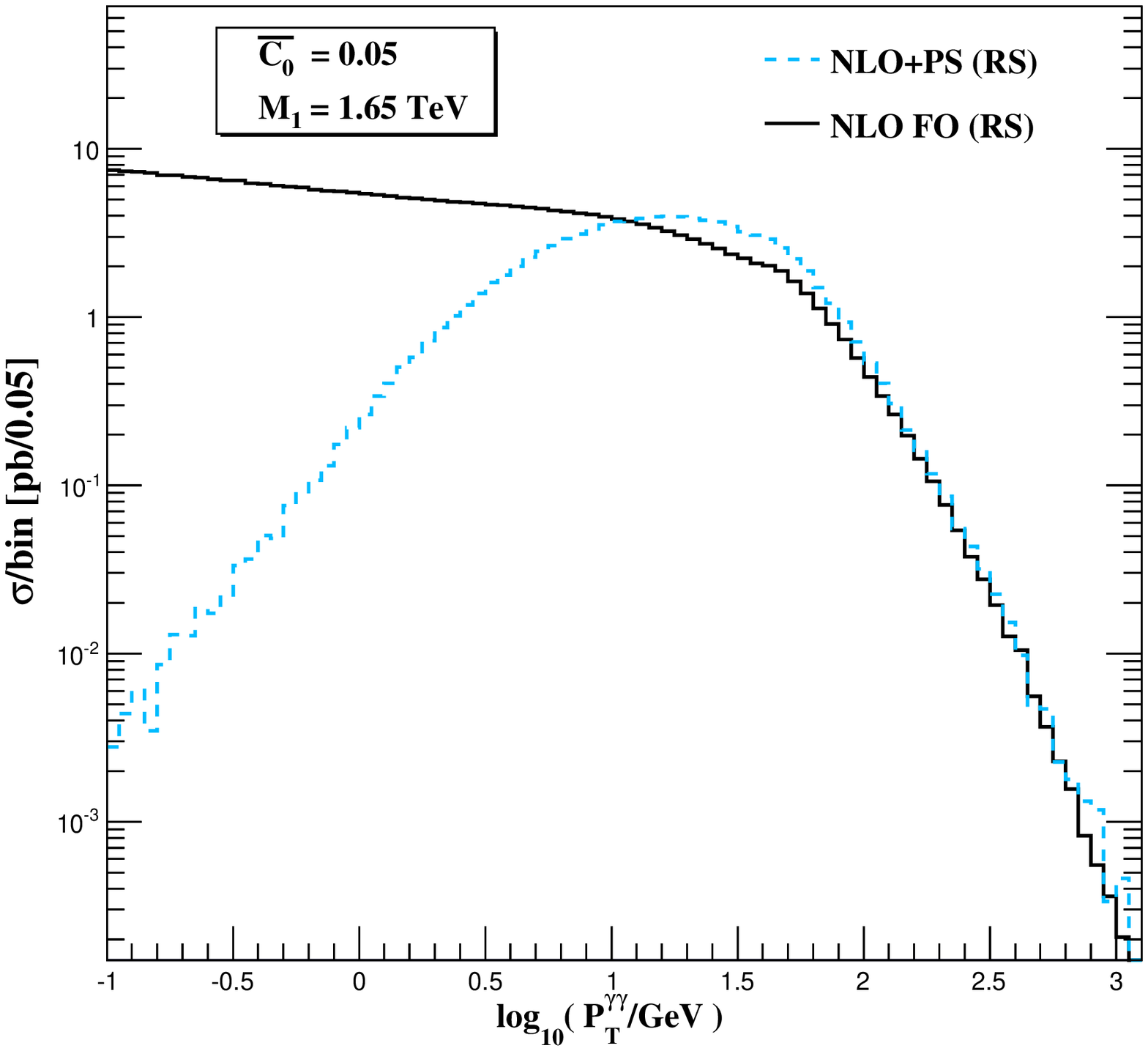}
 }
 \caption{Transverse momentum distribution of RS contribution, shown in log-log scale at fixed order NLO and NLO+PS for the Drell-Yan (left) and di-photon 
 (right) production processes.}
 \label{logpt_ee_aa}
\end{figure}

The events thus generated are then matched with HERWIG6 \cite{Corcella:2000bw}
parton shower Monte Carlo using the MC@NLO formalism \cite{Frixione:2002ik}.
While showering di-lepton events, $P_T^l \geq 20$ GeV, 
$|y^l|\leq2.5$, $\Delta R^{ll} > 0.4$ have been used for the analysis purpose,
where $l=e^+,e^-$. In order to separate leptons from jets $\Delta R^{lj} > 0.7$
has also been applied at this stage and finally we have found out hardest
$e^+$ and $e^-$ to build several kinematical observables with them. In case of
showering di-photon events, following analysis cuts are put on each photon with
the following FI parameters:  $P_T^{\gamma} \geq 20$ GeV, $|y^{\gamma}|\leq2.5$,
$\Delta R^{\gamma\gamma} > 0.4$ and $\epsilon_\gamma=1$, $r_0=0.4$, $n=2$ and
we have collected two hardest photons $\gamma_1$, $\gamma_2$ among many others.
As described earlier, Z bosons are leptonically decayed at the time of
showering $ZZ$ events and the applied analysis cuts are as follows:
$P_T^l \geq 20$ GeV, $|y^l|\leq2.5$, $\Delta R^{ll} > 0.4$,
$\Delta R^{lj} > 0.7$, where $l=e^+,e^-,\mu^+,\mu^-$. After that, all the
final state stable leptons ({\it i.e.}, $e^{\pm}, \mu^{\pm}$) are being
collected to make pair of leptons that have equal flavor but opposite charge.
Finally we have selected those leptons that are contributing as the hardest
$e^+e^-$ and $\mu^+\mu^-$ pairs with the condition that their invariant
masses ($M^{l^+l^-}$) satisfy the criteria $|M^{l^+l^-} - M_Z| < 10$ GeV, to
make sure that those leptons are actually decay products of the Z bosons. At
the time of showering $W^+W^-$ events with their corresponding decay modes,
we have identified the final state stable lepton-neutrino pair whose mother
is one of the $W^{\pm}$ bosons and make them pass through the following set
of analysis cuts: $P_T^l \geq 20$ GeV, $|y^l|\leq2.5$, $\Delta R^{ll} > 0.4$,
$\Delta R^{lj} > 0.7$ and $\cancel{E}_T > 30$ GeV. We have checked that all
the above events generated in each processes produce completely unbiased
results with the appliance of our present choices of generation and analysis
cuts.  
\begin{figure}
 \centering{
 \includegraphics[height=8.0cm,width=7.4cm]{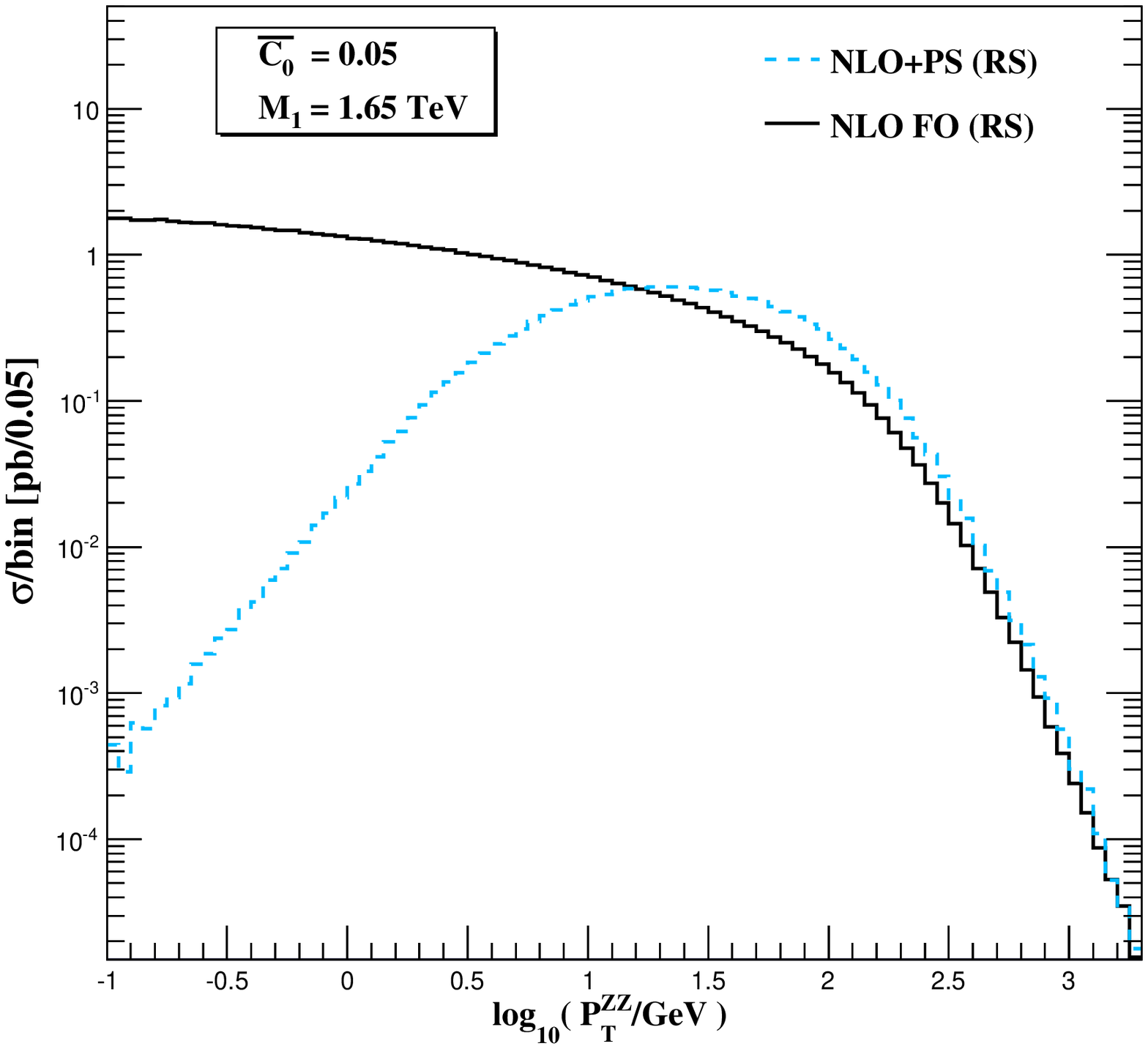}
 \includegraphics[height=8.0cm,width=7.4cm]{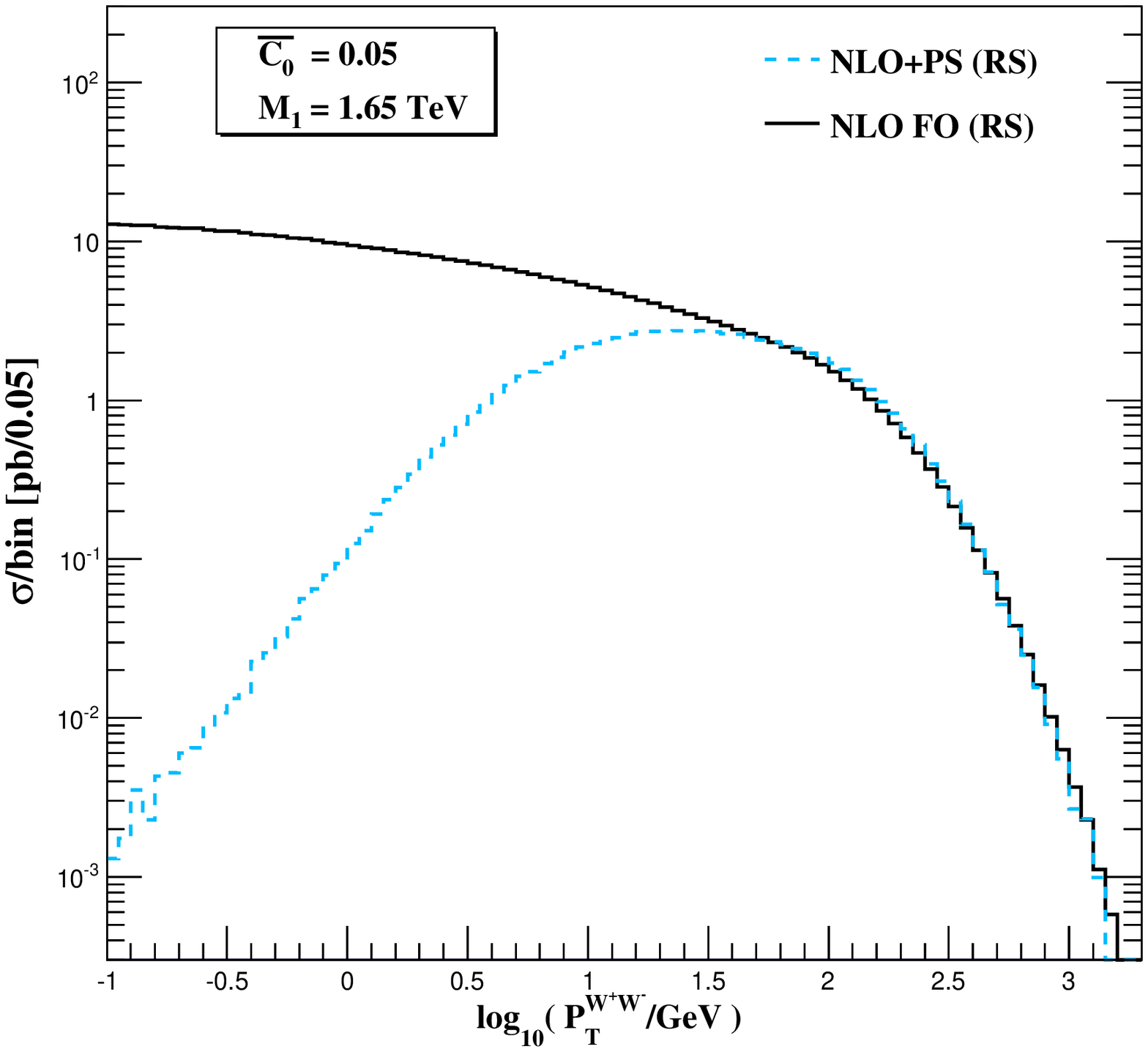}
 }
 \caption{Transverse momentum distribution of RS contribution, shown in log-log scale at fixed order NLO and NLO+PS for the $ZZ$ (left) and $W^+W^-$ 
 (right) production processes.}
  \label{logpt_zz_ww}
\end{figure}

To show the effect of parton shower over the fixed order NLO results, we
have presented in Fig.\ (\ref{logpt_ee_aa}) the $\log_{10}P_T$ distributions 
of the $e^+e^-$ (left) and $\gamma\gamma$ (right) pair at fixed order 
NLO (solid black) as well as in NLO+PS (dashed blue) accuracy for the RS
case.  Both the curves in each figure are plotted using respective
analysis cuts and we have used the RS model parameter $\overline{c_0}=0.05$
and the corresponding $M_1$ value is taken as $M_1=1.65$ TeV.  Note that, by the label
\textquoteleft RS\textquoteright\ in the figure, we mean the total
contribution that consists of SM, RS and the interference between
the two and we maintain the same convention in the rest of the figures 
as well. Likewise, Fig.\ (\ref{logpt_zz_ww}) represents similar
distributions for the $ZZ$ (left) and $W^+W^-$ (right) pairs. Each of these
figures shows diverging nature of the fixed order NLO curve in the
$P_T\rightarrow 0$ region, whereas the NLO+PS result gets convergent
in that region ensuring the correct resummation of the Sudakov
logarithms and thereby leading to a suppression in cross section
in the low-$P_T$ region.  As expected, both the results are in good
agreement with one another in the high-$P_T$ region. 
%
%%%%% DY plots begins %%%%%
\begin{figure}
 \centering{
 \includegraphics[height=9.5cm,width=12cm]{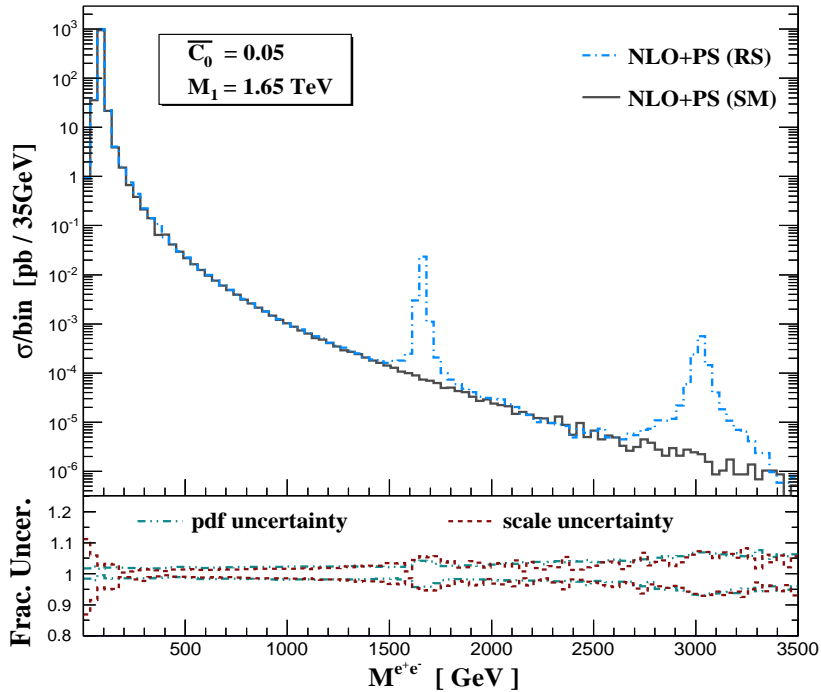}
 }
 \caption{Invariant mass distribution of the di-lepton pair in Drell-Yan
  process for RS and SM.}
 \label{EE_ee_inv}
\end{figure}

Uncertainty calculations of various distributions have been performed
automatically in the {\sc a}MC@NLO framework by using its built-in
re-weighting procedure that stores sufficient information in the
parton level Les Houches event files. Independent variation of $\mu_R$
and $\mu_F$ scales are considered to calculate scale uncertainties. We
have set $\mu_R=\xi_R M$ and $\mu_F=\xi_F M$, where M denotes the
invariant mass of the di-final state ({\it i.e.}, $M^{e^+e^-}$,
$M^{\gamma\gamma}$, $M^{ZZ}$ or $M^{W^+W^-}$, as applicable) and $\xi_R$,
$\xi_F$ can take any one of the values ($1,1/2,2$) at a time. The scale
uncertainty band would then be determined as the envelope
\cite{Frederix:2012dp} of the following ($\xi_R, \xi_F$) combinations: 
($1,1$), ($1,1/2$), ($1,2$), ($1/2,1/2$), ($1/2,1$), ($2,1$), ($2,2$).
On the other hand, PDF uncertainties are estimated using Hessian method
as prescribed by the MSTW collaboration \cite{Martin:2009iq}. 
%
%%%%% DY plots %%%%%
\begin{figure}
 \centering{
 \includegraphics[height=8.5cm,width=7.4cm]{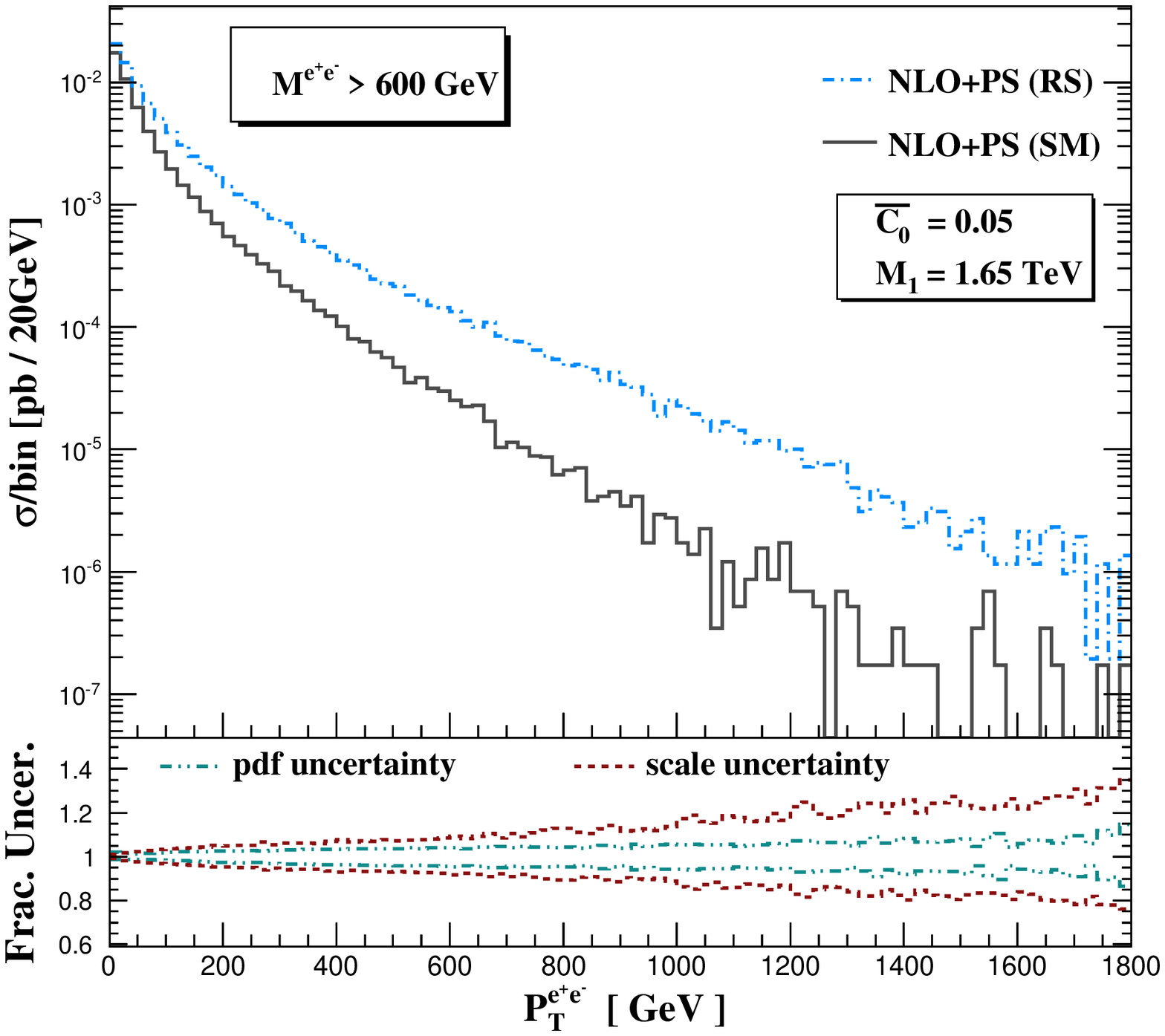}
 \includegraphics[height=8.5cm,width=7.4cm]{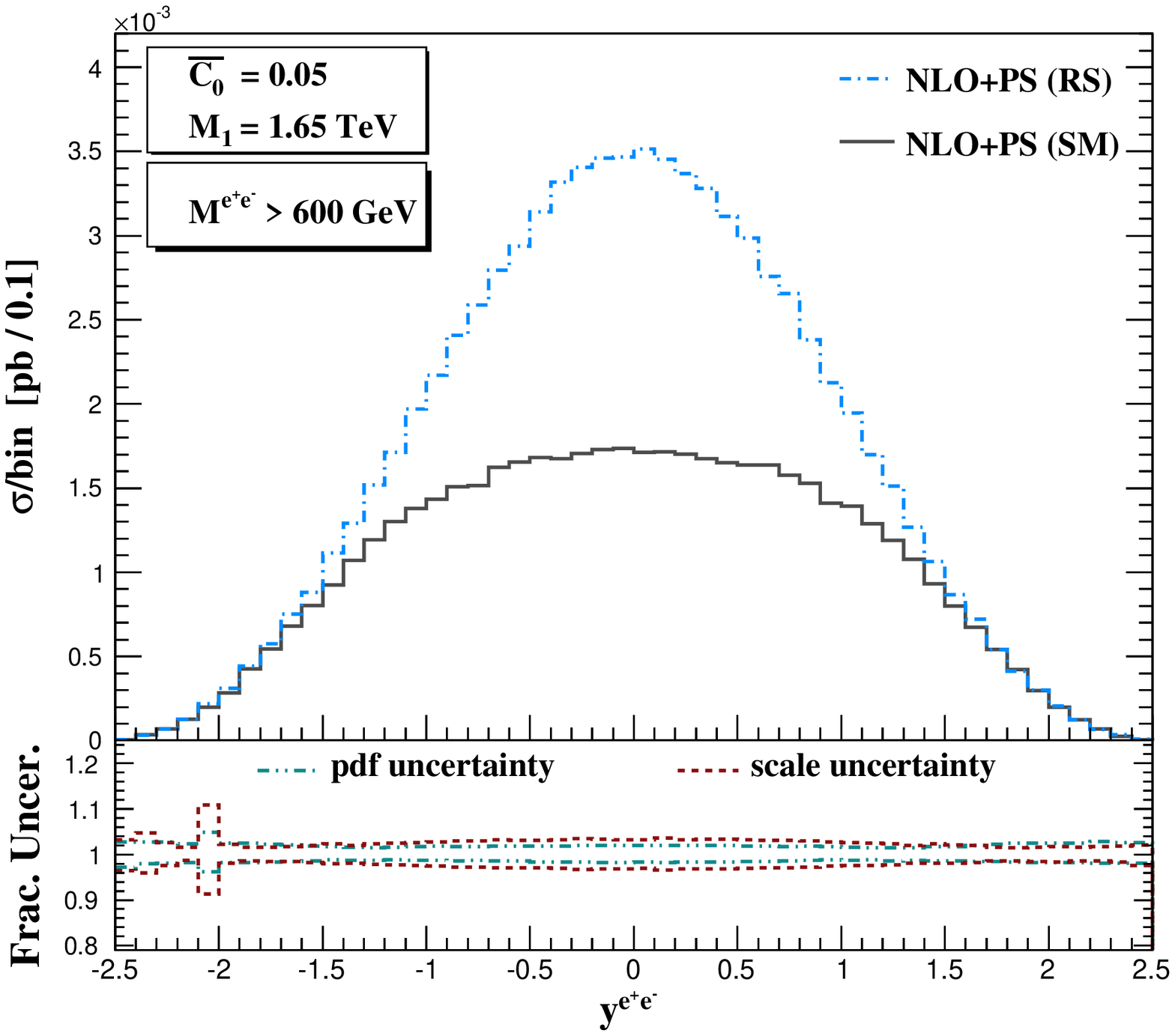}
 }
 \caption{Transverse momentum (left) and rapidity (right) distribution
 of the di-lepton pair in both RS and SM.}
 \label{EE_ee_pt_rap}
\end{figure}

In all the subsequent figures, various distributions of kinematical
observables are depicted using a consistent graphical representation. 
In the Figures the main frame, depicts distributions that are of 
outcome of both the RS
(dash-dotted blue) and SM (solid black), are shown to NLO+PS accuracy,
whereas the corresponding lower insets provide the estimation of the
fractional scale (dashed red) and PDF (dash-double dotted green) uncertainties,
which basically denote the variation of the central value ({\it i.e},
the extremum value divided by the central value). Unless stated
otherwise, $\overline{c_0}=0.05$ and $M_1=1.65$ TeV are used in all these
plots. 
%
%%%%% DY plots %%%%%
\begin{figure}
 \centering{
 \includegraphics[height=8.5cm,width=7.4cm]{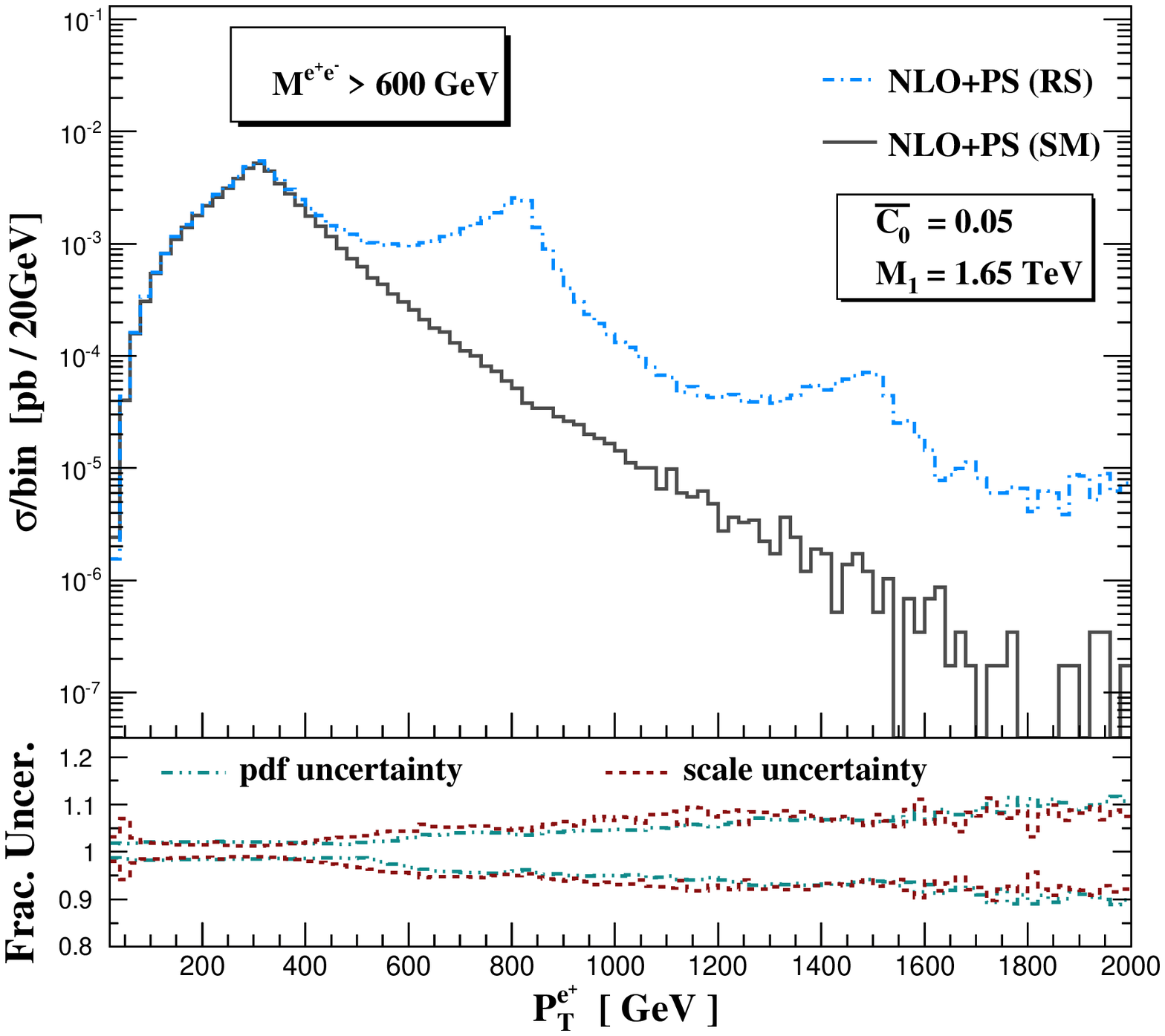}
 \includegraphics[height=8.5cm,width=7.4cm]{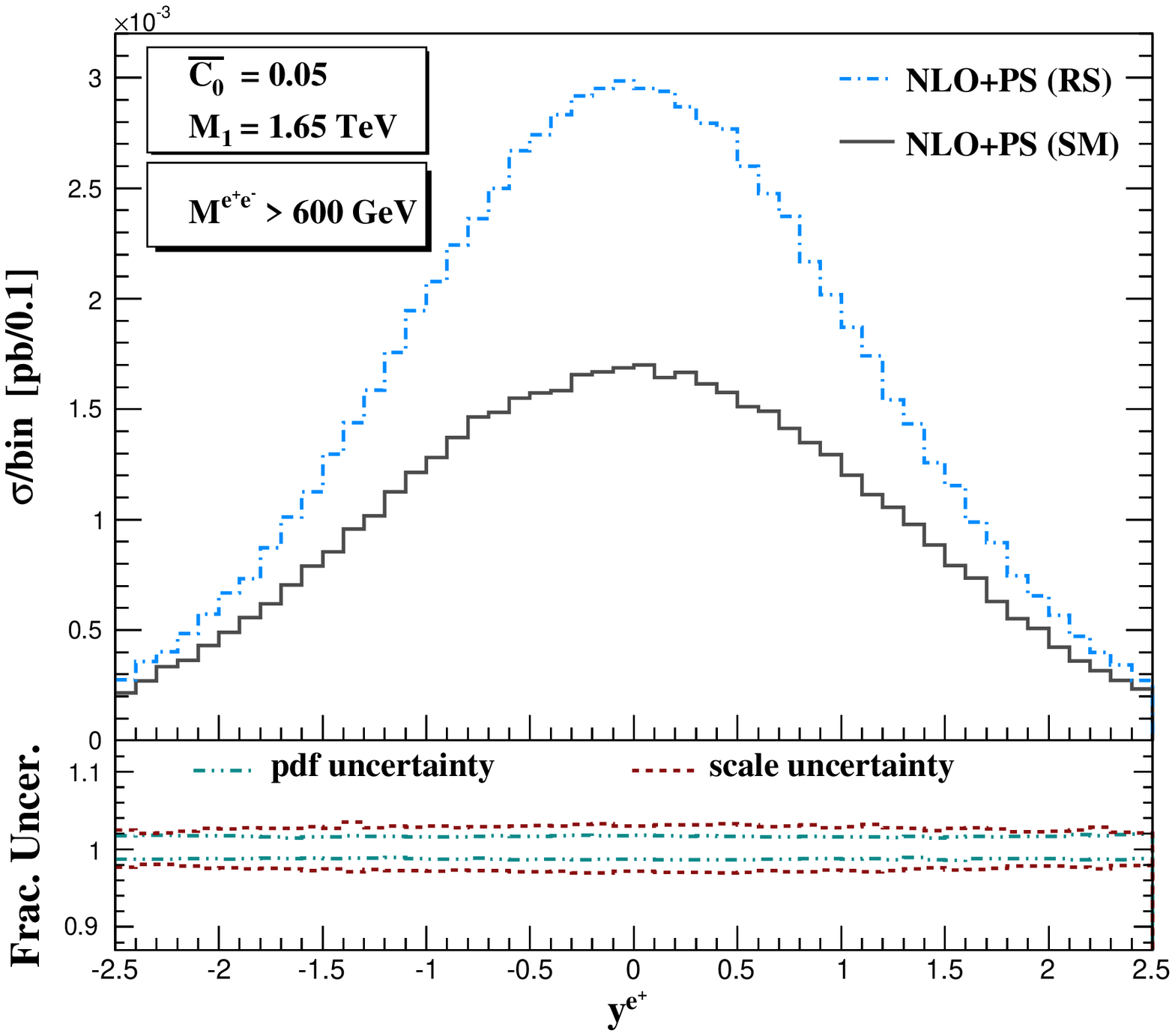}
 }
 \caption{Transverse momentum (left) and rapidity (right) distribution
 of the positron for the Drell-Yan production process in SM and RS.}
 \label{EE_ep_pt_rap}
\end{figure}

\begin{figure}
 \centering{
 \includegraphics[height=8.5cm,width=7.4cm]{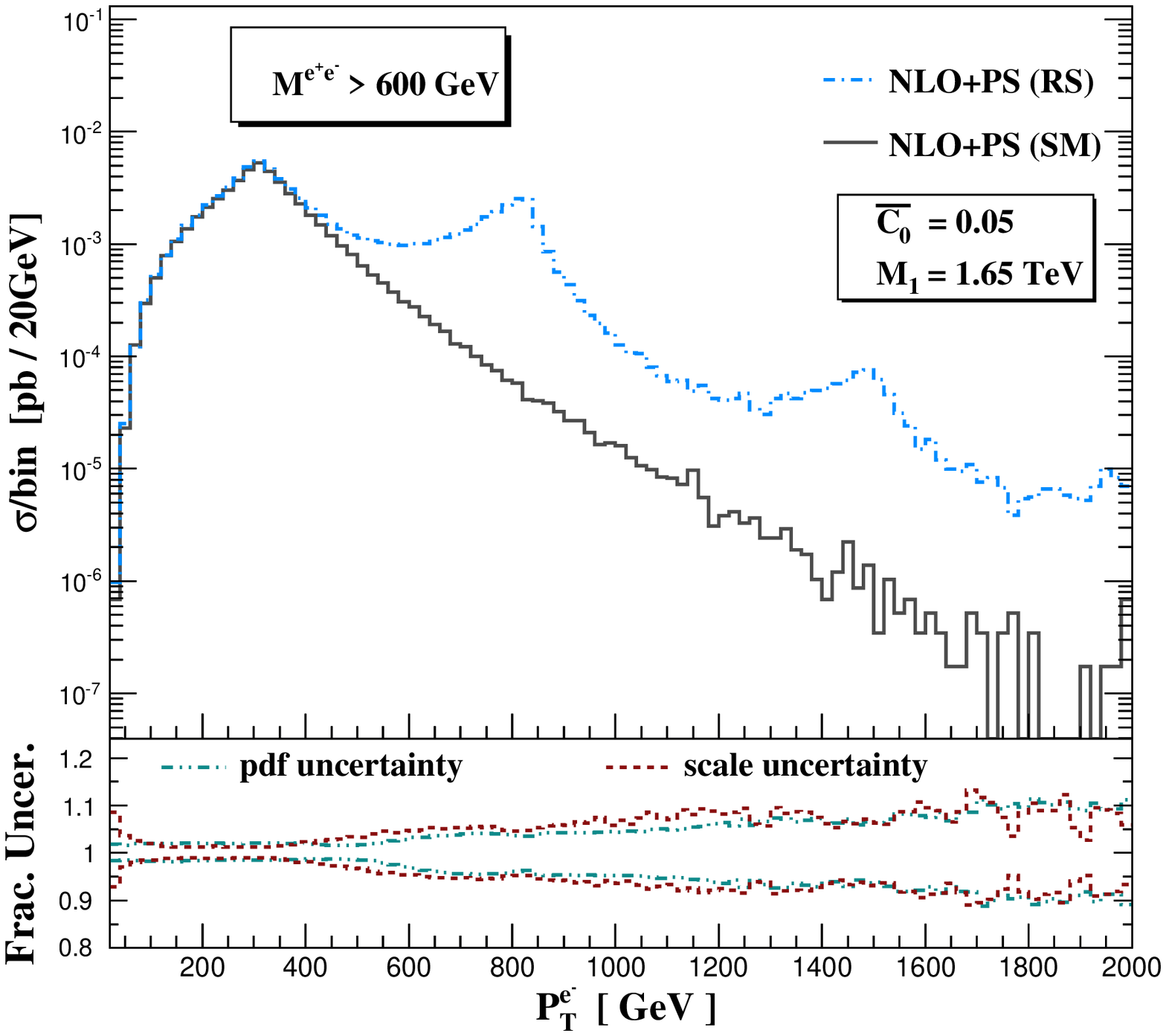}
 \includegraphics[height=8.5cm,width=7.4cm]{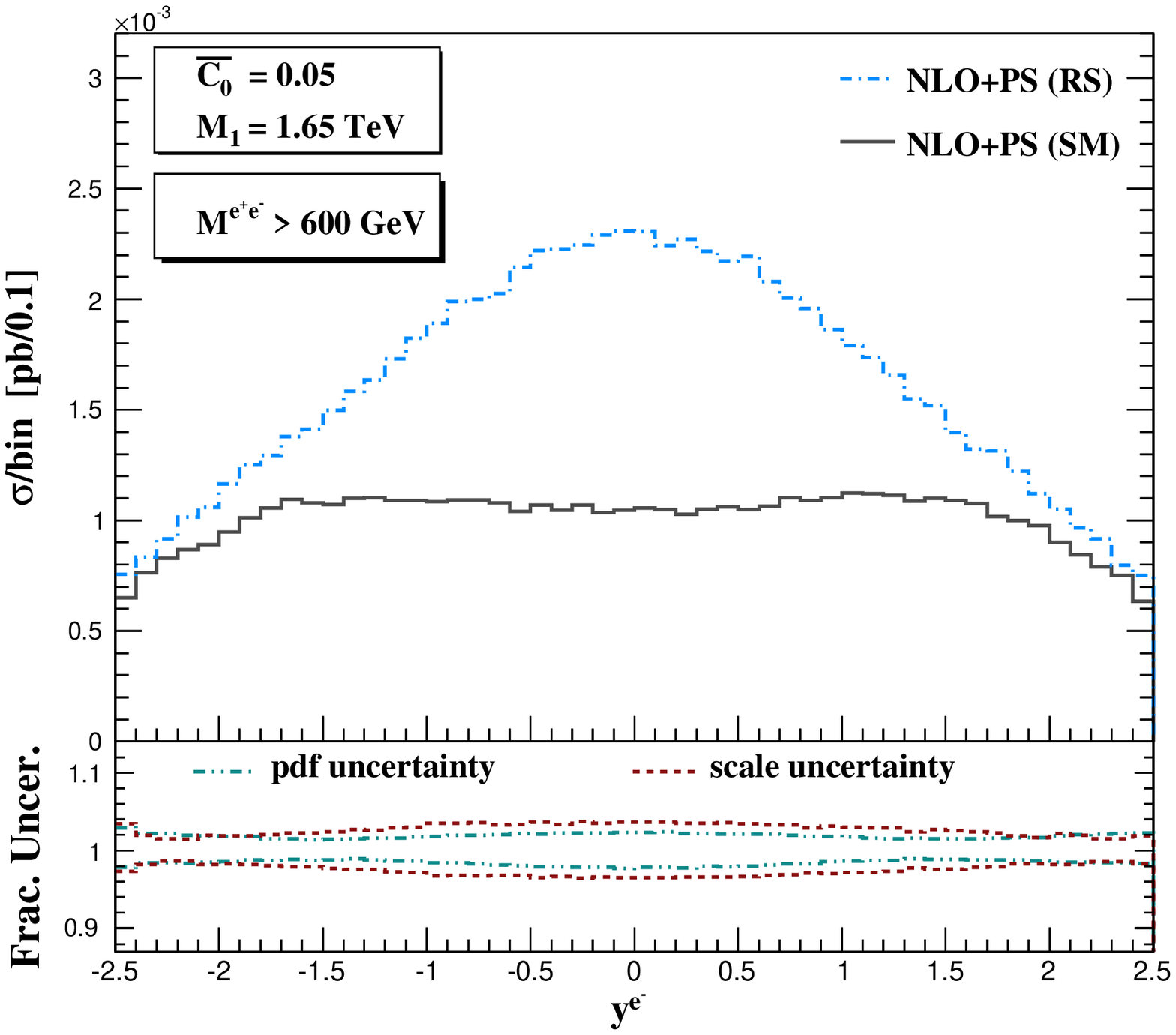}
 }
 \caption{Transverse momentum (left) and rapidity (right) distribution of the 
electron for the Drell-Yan production process in SM and RS.}
 \label{EE_em_pt_rap}
\end{figure}

For the Drell-Yan production process, invariant mass distribution
of the $e^+e^-$ pair is shown in Fig.\ (\ref{EE_ee_inv}).  The two
peaks in the RS case indicate first ($M_1$) and second ($M_2$)
excitations of the RS graviton and they perfectly match with
the theoretical values (see Eq. \ref{gmass}).  Fig.\ (\ref{EE_ee_pt_rap}),
(\ref{EE_ep_pt_rap}) and (\ref{EE_em_pt_rap}) apprise the transverse
momentum (left) and rapidity (right) distributions of the $e^+e^-$
pair, $e^+$ and $e^-$ respectively in the region $M^{e^+e^-}>600$
GeV.  Such invariant mass cut, which is also applied consistently
to the rest processes, is an optimal choice to reduce SM background
effects, ensuring the influence of sufficient signal events at the same
time.  Individually, transverse momentum distributions of $e^+$ and
$e^-$ are of similar kind and two kinks are arising near the half of
the $M_1$ and $M_2$ values. However, in the combined transverse
momentum distribution of the $e^+e^-$ pair, washing out of such
kinks points out that the directions of outgoing $e^+$ and $e^-$
are opposite to each other in the transverse plane of the beam direction. 
The fractional uncertainties associated with these $P_T$ distributions
are large in the high transverse momentum region, although they are
quite minimal in the low-$P_T$ region, where the higher order effects
are included as a result of resummation.  In the high-$P_T$ region
of $e^+ e^-$ pair, the large uncertainty is a reflection of the fact
that it is in fact a leading order process.  
Note that, the rapidity distributions of $e^+$ and
$e^-$ are not similar because the high invariant mass cut is
responsible in breaking the angular correlation between them. To
estimate the improvement in the results while including NLO corrections, 
we find that the scale uncertainties in the central rapidity region
of all these rapidity distributions in the RS case reduce
to about 6-8\% in NLO+PS, from about 10-12\% at LO+PS. PDF uncertainties
in NLO+PS are about 1-1.5\% less compared to the LO+PS outcomes. 
%
%%%%% DY plots %%%%%
%%%%% DY plots ends %%%%%

%%%%% Di-photon plots begins %%%%%
\begin{figure}
 \centering{
 \includegraphics[height=8.5cm,width=7.4cm]{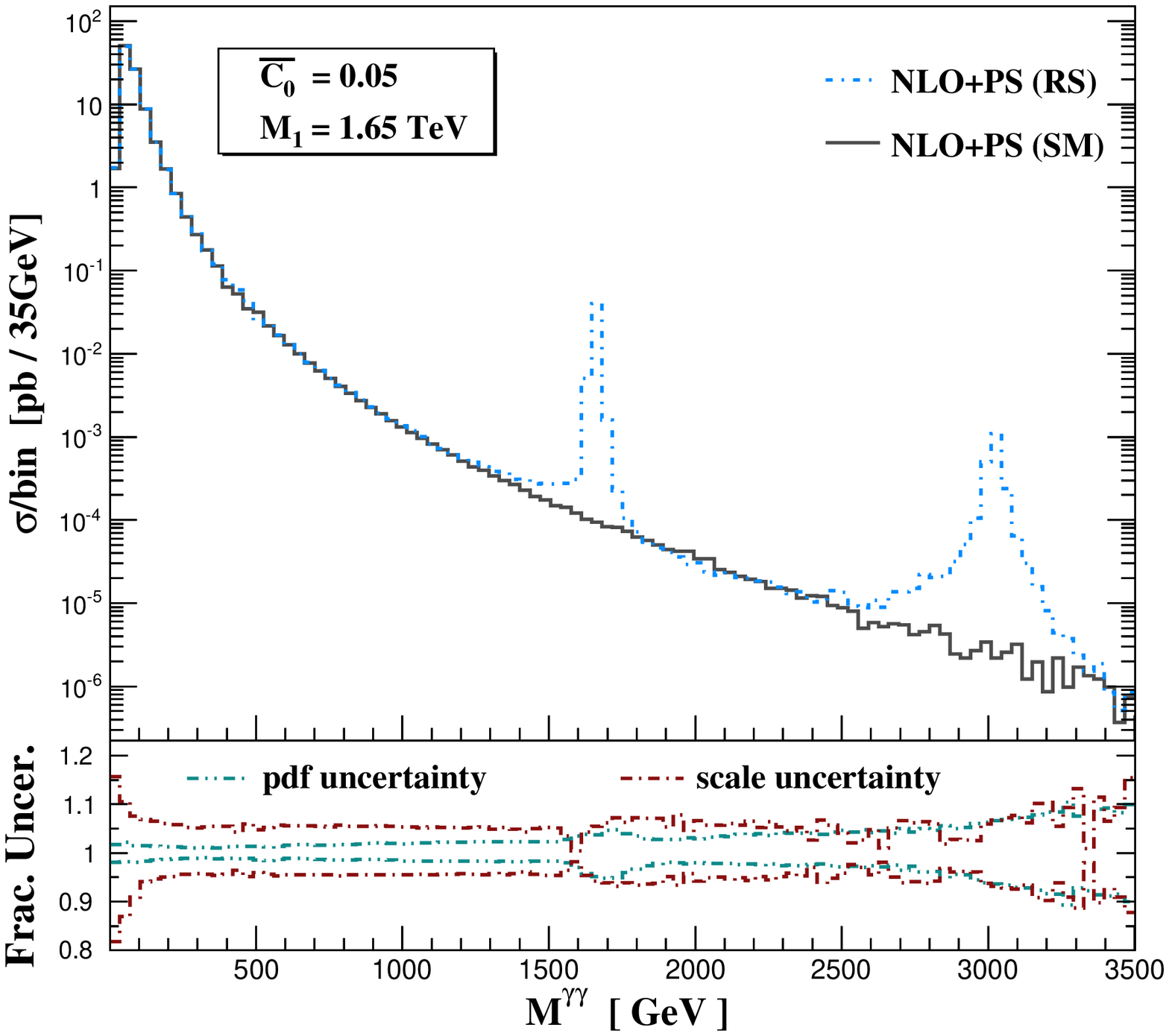}
 \includegraphics[height=8.5cm,width=7.4cm]{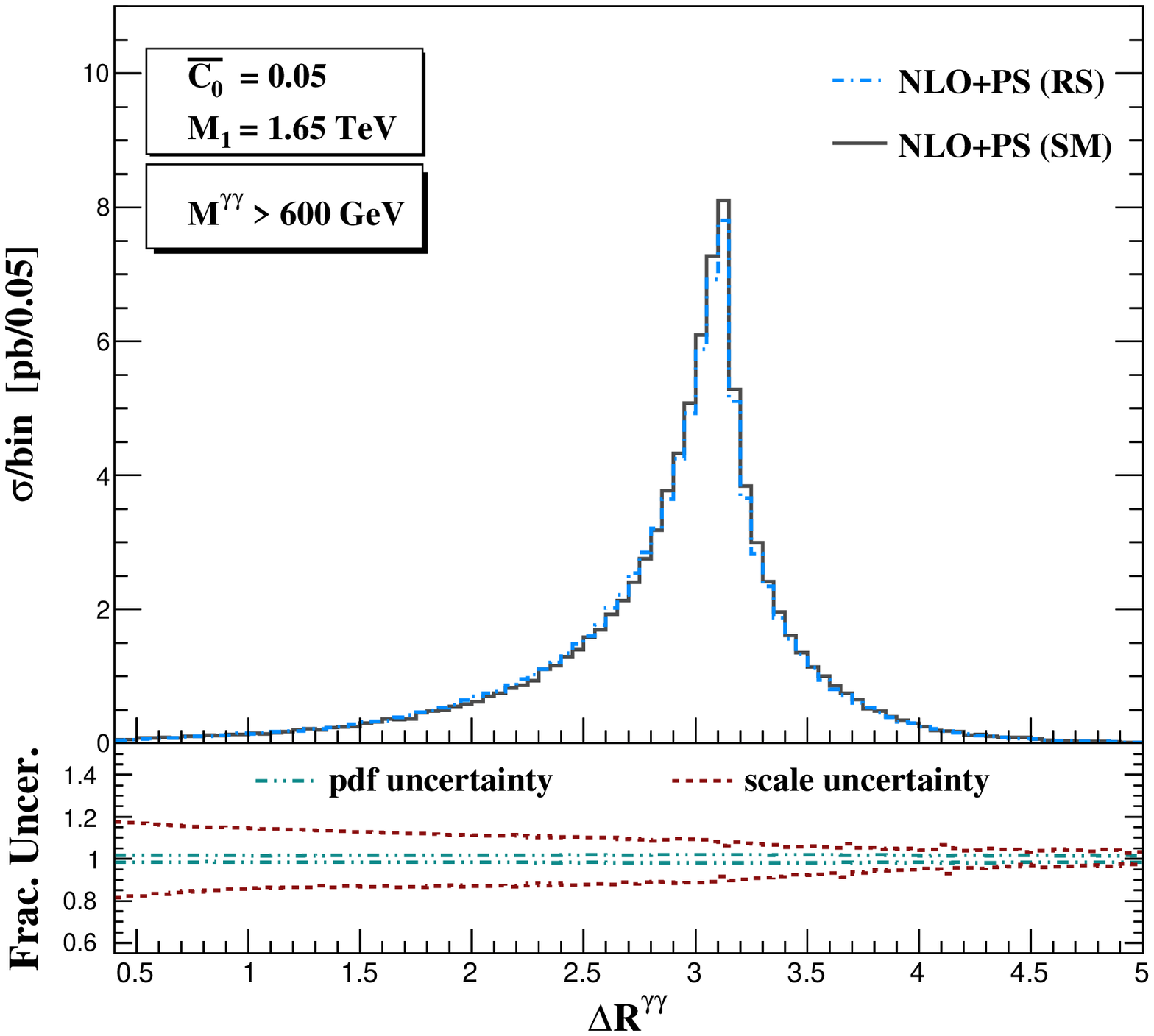}
 }
 \caption{Invariant mass distribution of the di-photon pair (left) in RS and SM. The right panel shows the separation between two 
 hardest photons in the rapidity-azimuthal angle plane.}
 \label{AA_aa_inv_dr}
\end{figure}

\begin{figure}
 \centering{
 \includegraphics[height=8.5cm,width=7.4cm]{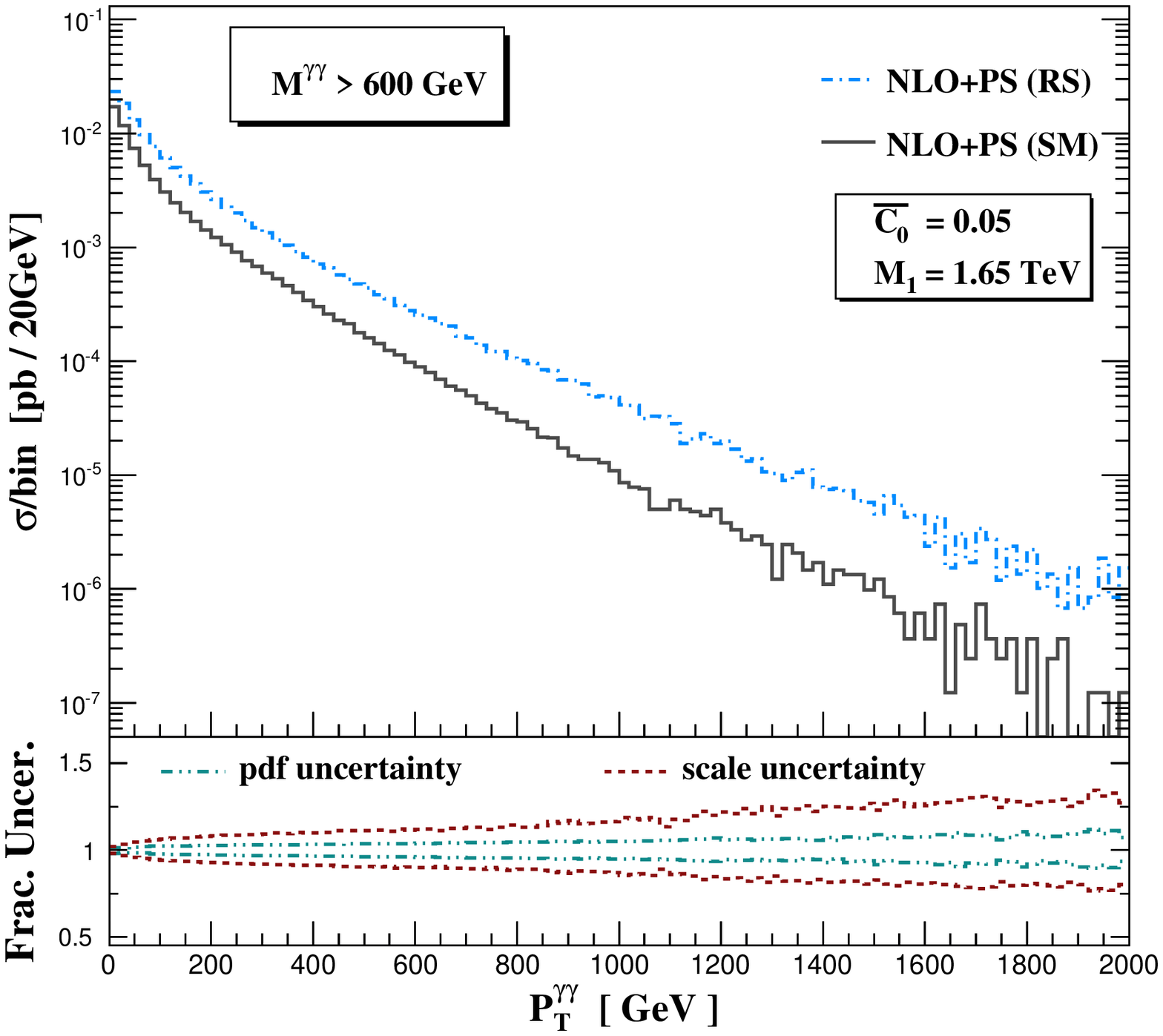}
 \includegraphics[height=8.5cm,width=7.4cm]{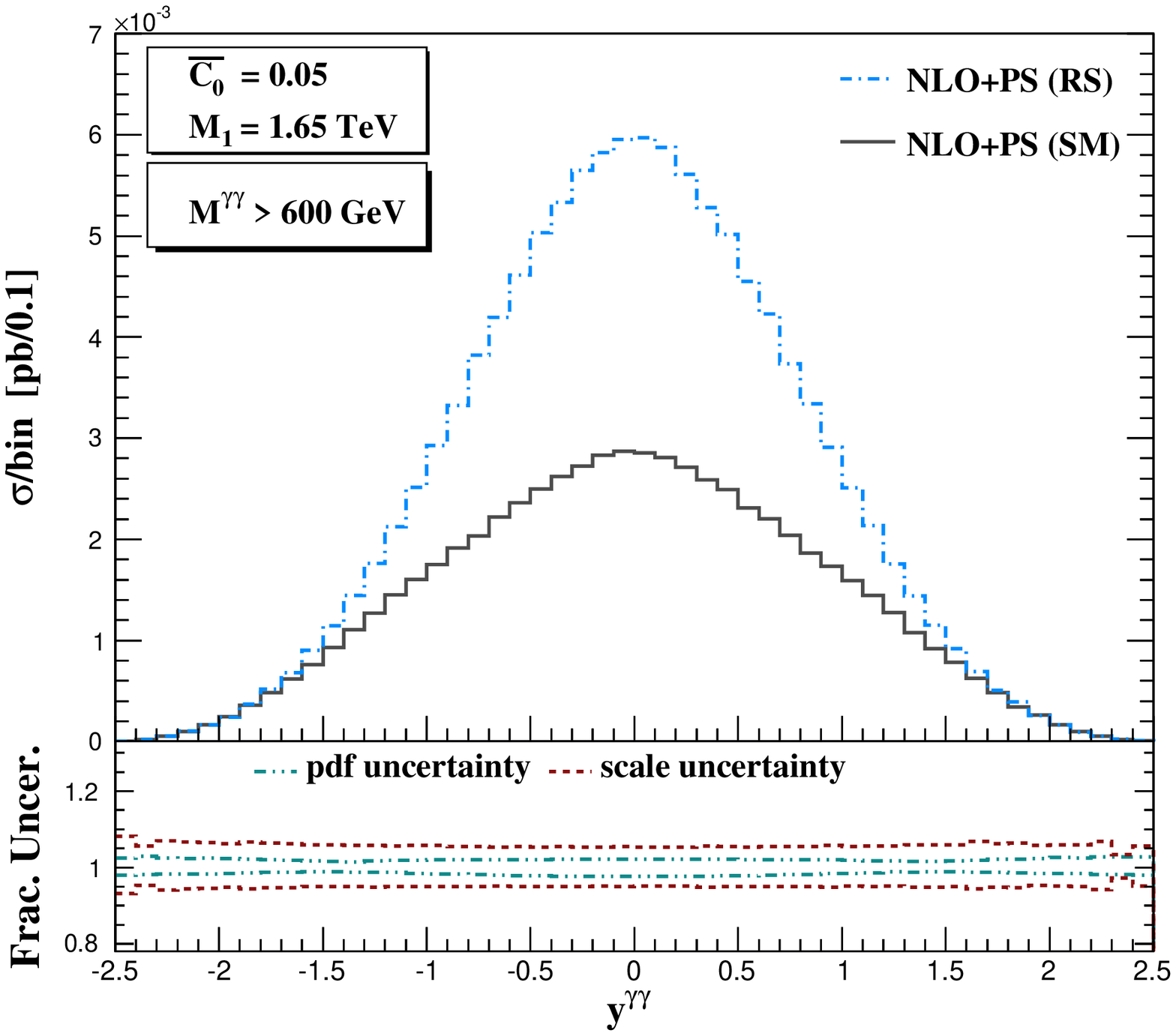}
 }
 \caption{Di-photon transverse momentum (left) and rapidity (right) distribution in RS and SM.}
 \label{AA_aa_pt_rap}
\end{figure}

\begin{figure}
 \centering{
 \includegraphics[height=8.5cm,width=7.4cm]{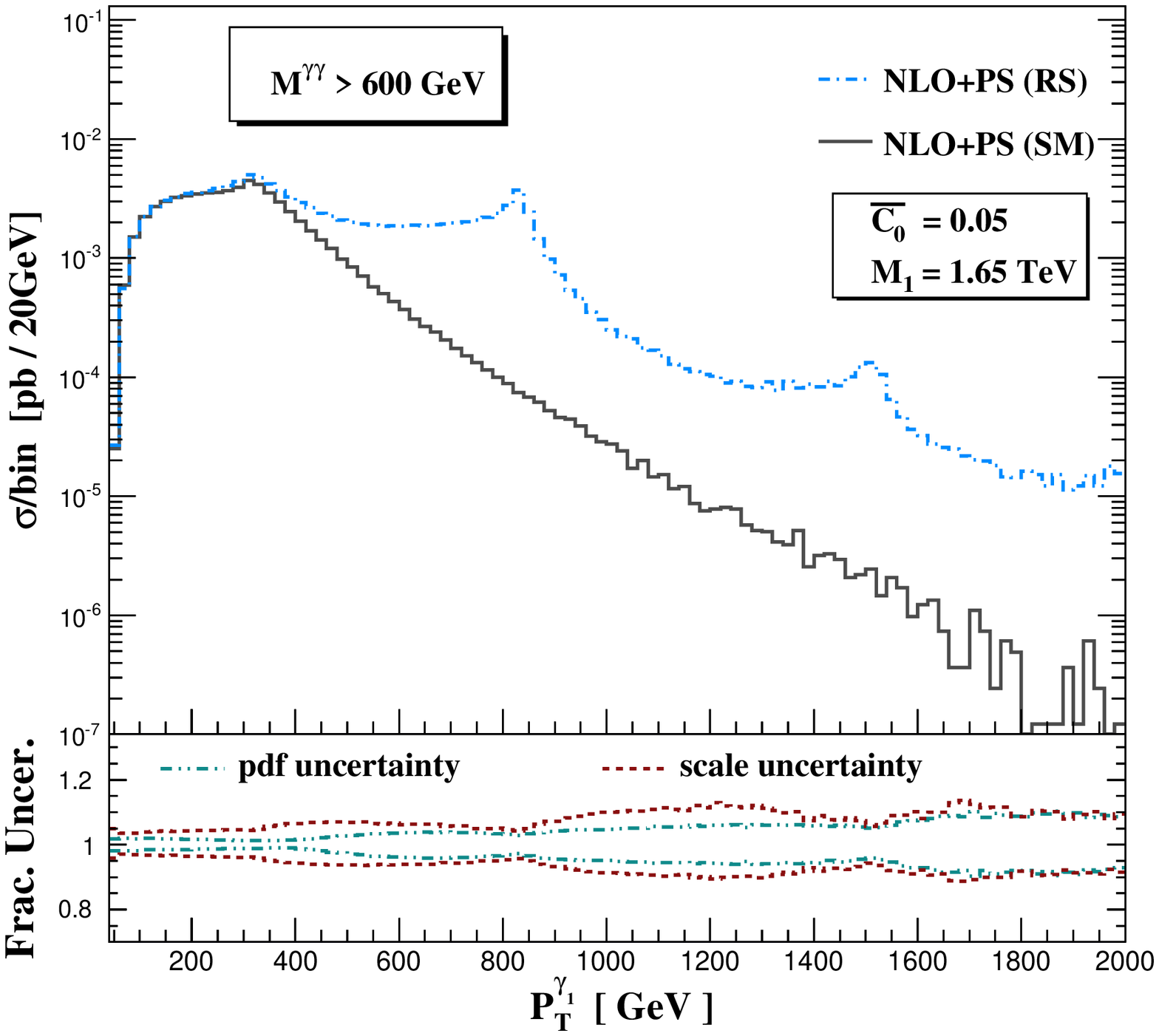}
 \includegraphics[height=8.5cm,width=7.4cm]{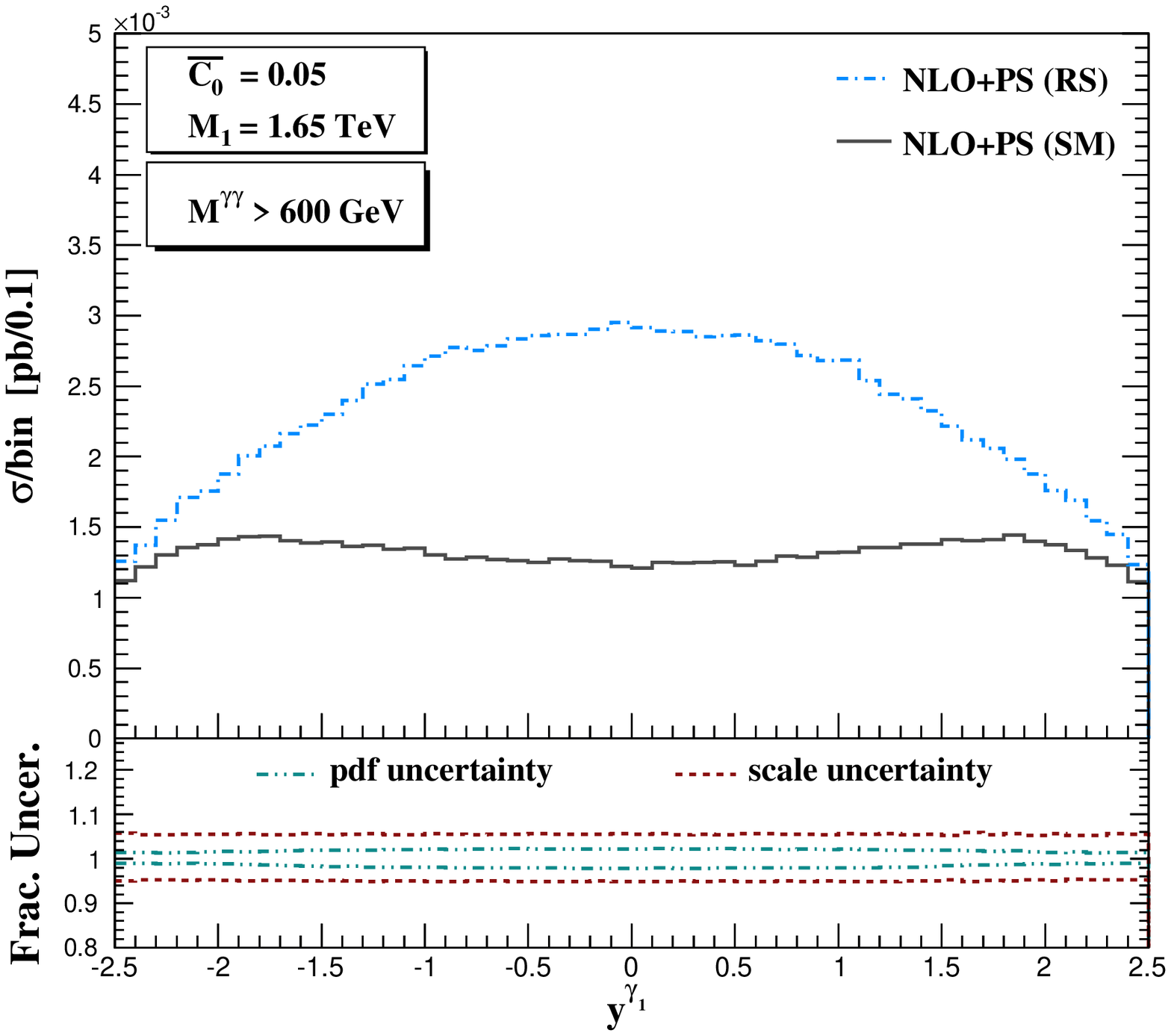}
 }
 \caption{Transverse momentum (left) and rapidity (right) distribution of the hardest photon in di-photon production process 
 in RS and SM.}
 \label{AA_a1_pt_rap}
\end{figure}

Distributions related to the di-photon production are shown in Fig.\ (\ref{AA_aa_inv_dr})-(\ref{AA_a1_pt_rap}). In Fig.\ (\ref{AA_aa_inv_dr}), we have 
presented invariant mass (left) distribution of the diphoton system and the separation (right) between the two hardest photons in the rapidity-azimuthal 
angle plane with the cut $M^{\gamma\gamma}>600$ GeV. The invariant mass distribution clearly shows two peaks that correspond to the choice 
of $M_1$ value and the respective $M_2$ value derived from that. The $\Delta R^{\gamma\gamma}$ distributions are almost same for SM and RS and the 
peaks near the angle $\pi$ ({\it i.e.}, $180\;\mathring{}$\;) in these distributions indicate the abundance of such two hardest photons that 
are mostly back-to-back and the associated scale uncertainty is becoming almost nil as the two photons are getting much away from one another. 
Fig.\ (\ref{AA_aa_pt_rap}) and (\ref{AA_a1_pt_rap}) represent the transverse momentum (left) and rapidity (right) distributions of the di-photon 
system and the hardest photon respectively in the region where the condition $M^{\gamma\gamma}>600$ GeV is satisfied. Here also, we are getting 
two kinks in the transverse momentum distribution of the individual hardest photon around the half of the first and second excitation values of 
the RS graviton. The scale uncertainties in the central rapidity regions in LO+PS were around 13-14\% and they are as expected reduced to 
10\% in NLO+PS, although the reduction in PDF uncertainties is only about 0.2\% between the NLO+PS and LO+PS results. 

%%%%% ZZ plots begins %%%%%
\begin{figure}
 \centering{
 \includegraphics[height=9.5cm,width=12cm]{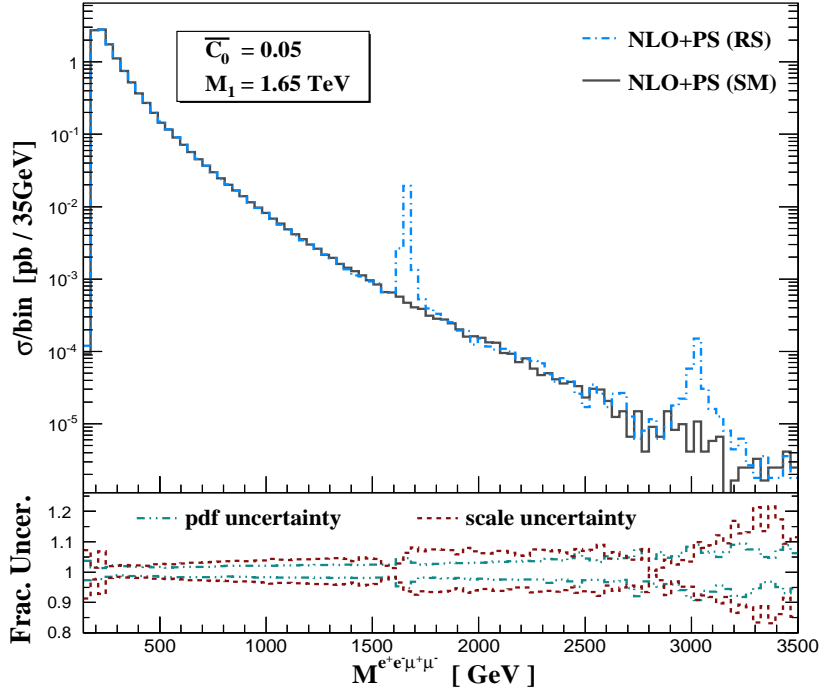}
 }
 \caption{Four-lepton invariant mass ($M^{e^+e^-\mu^+\mu^-}$) distribution for RS and SM coming from the decay products of $ZZ$ process.}
 \label{ZZ_2e2mu_inv}
\end{figure}

\begin{figure}
 \centering{
 \includegraphics[height=8.5cm,width=7.4cm]{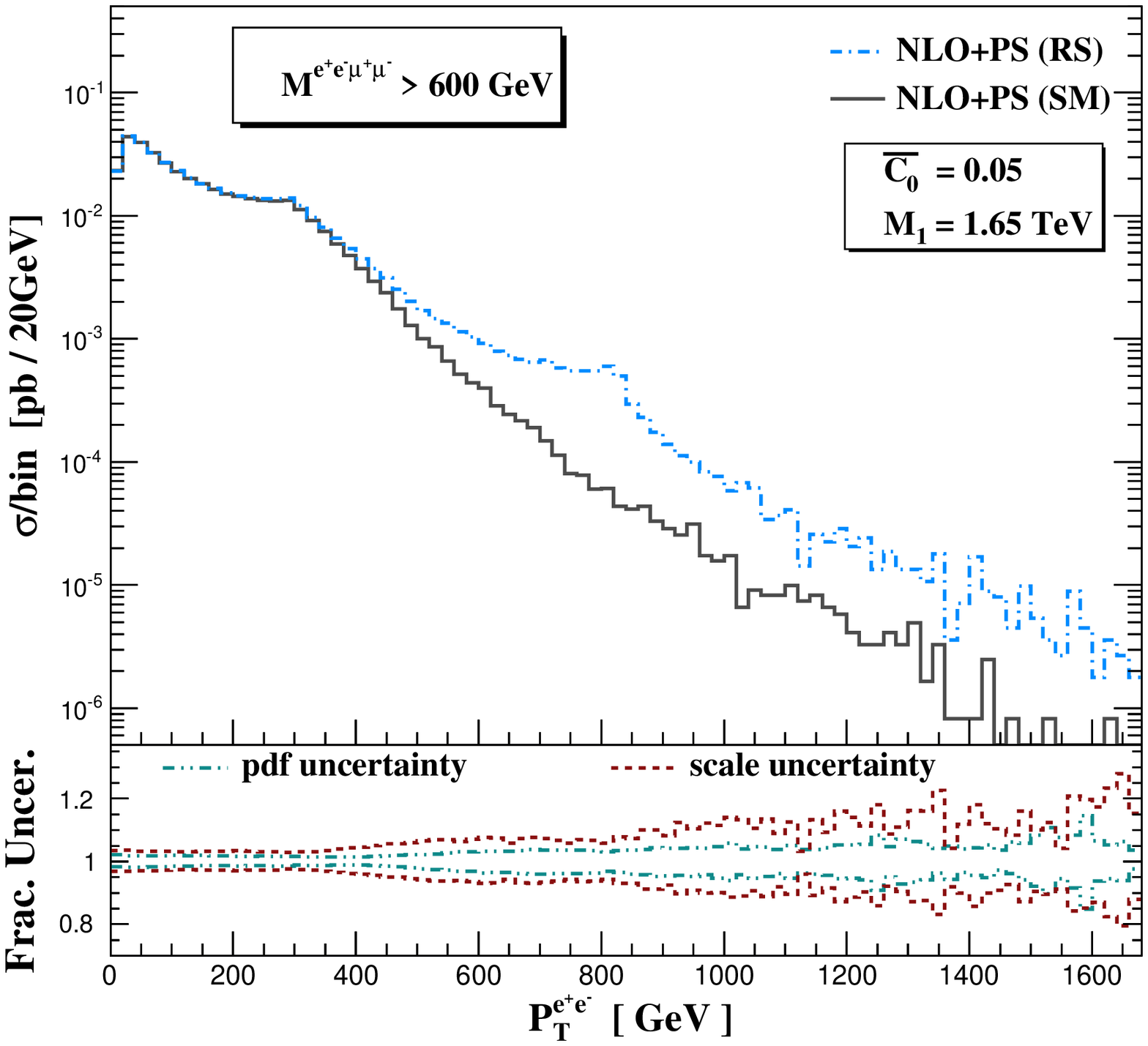}
 \includegraphics[height=8.5cm,width=7.4cm]{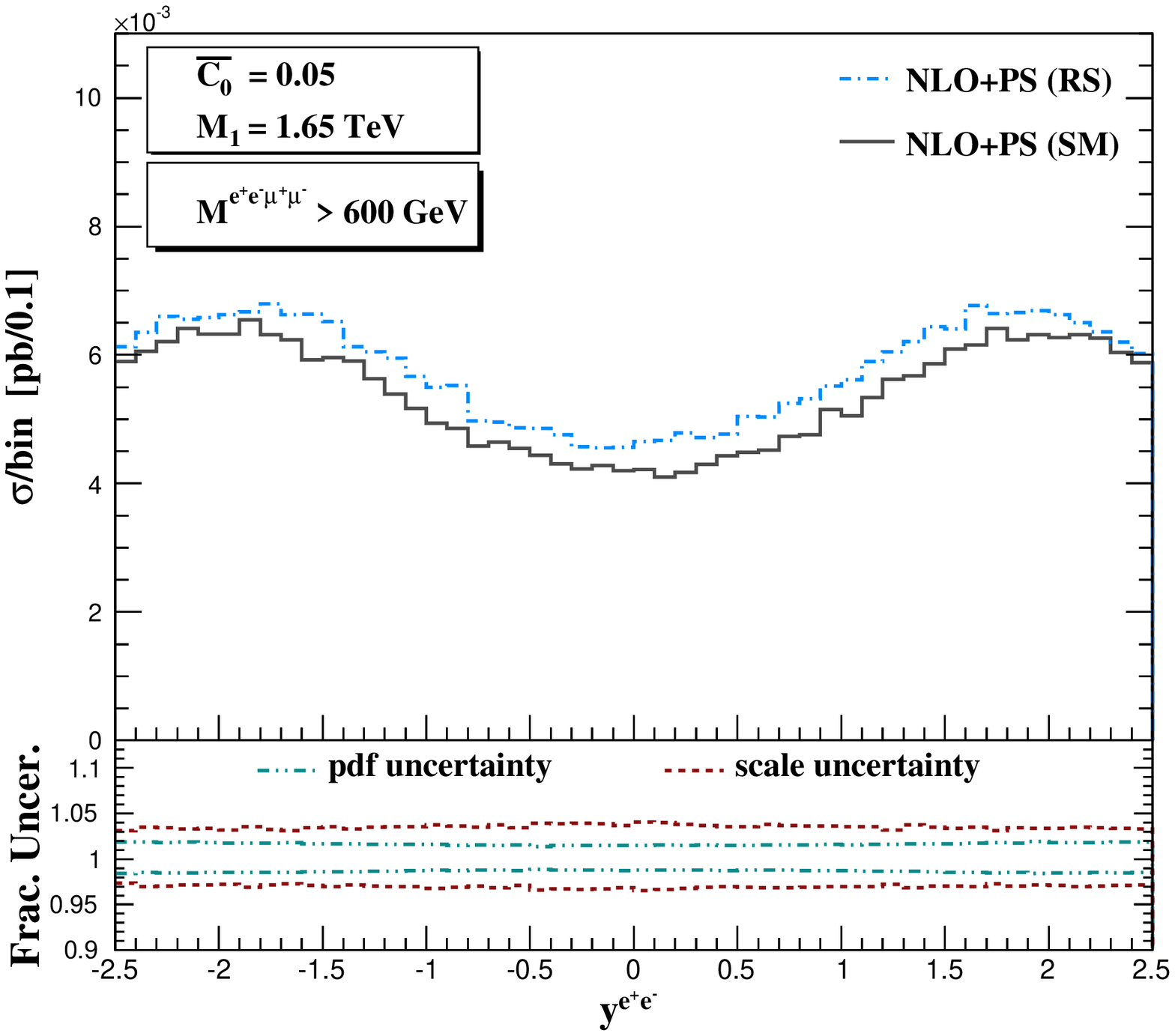}
 }
 \caption{Transverse momentum (left) and rapidity (right) distribution of
 the $e^{+} e^{-}$ pair coming from the decay products of ZZ process in RS and SM.}
 \label{ZZ_2e_pt_rap}
\end{figure}

\begin{figure}
 \centering{
 \includegraphics[height=8.5cm,width=7.4cm]{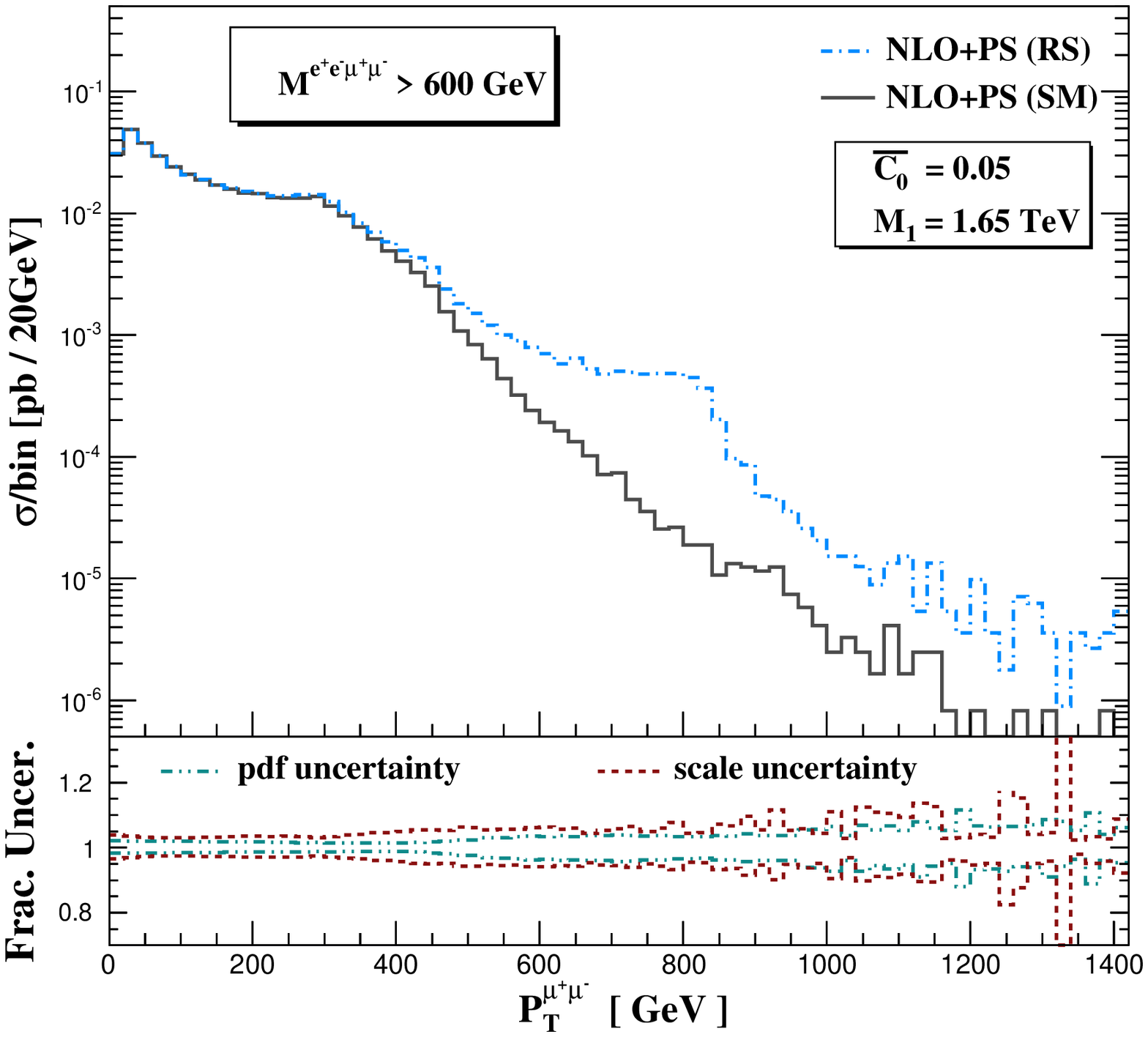}
 \includegraphics[height=8.5cm,width=7.4cm]{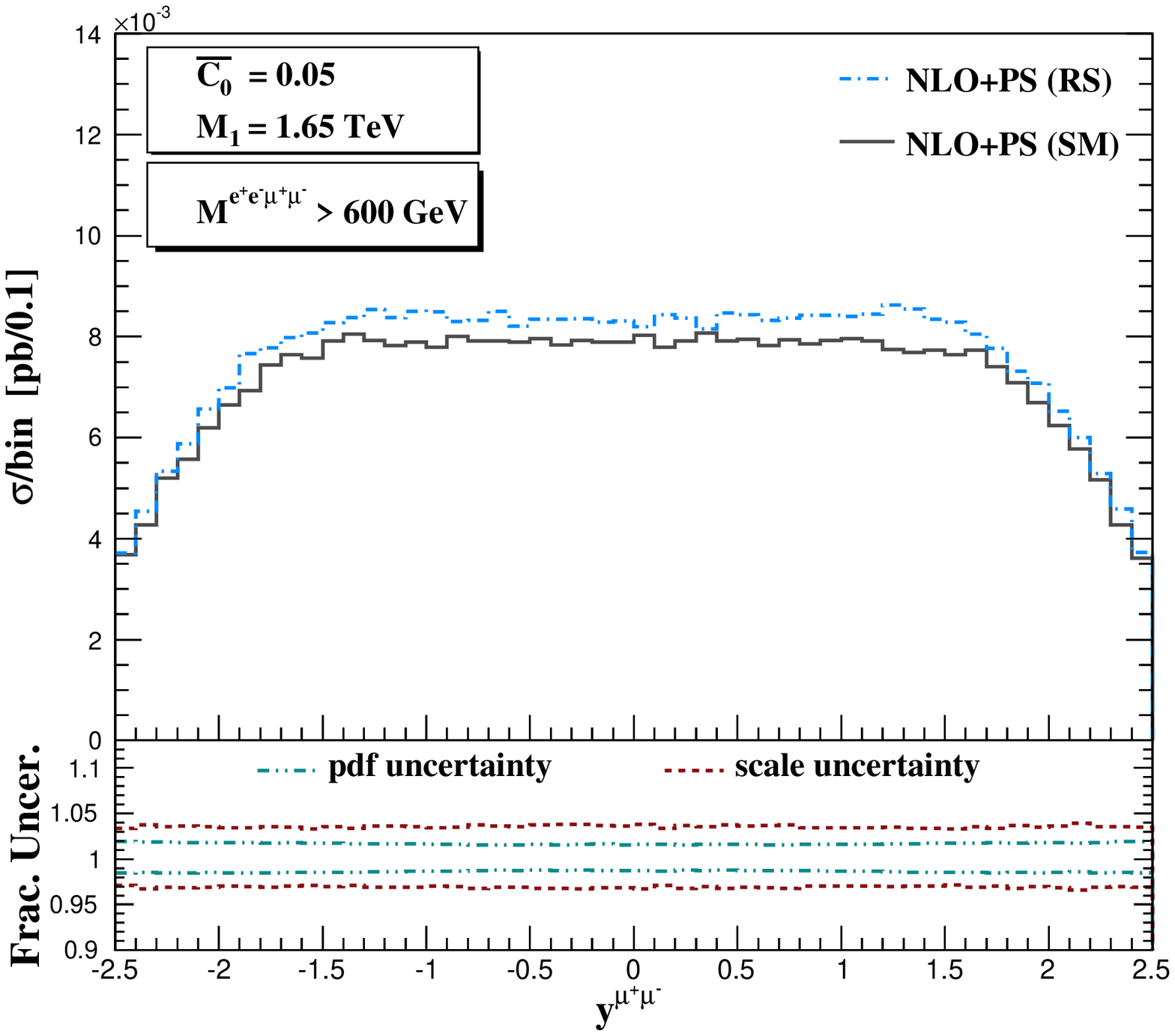}
 }
 \caption{Transverse momentum (left) and rapidity (right) distribution of
 the $\mu^+\mu^-$ pair coming from ZZ decay for RS and SM.}
 \label{ZZ_2mu_pt_rap}
\end{figure}

%%%%% ZZ plots ends %%%%%

Fig.\ (\ref{ZZ_2e2mu_inv})-(\ref{ZZ_2mu_pt_rap}) correspond to the
distributions of decay products that are coming from the $ZZ$ events.
The invariant mass ($M^{e^+e^-\mu^+\mu^-}$) distribution of all the
final state leptons is depicted in Fig.\ (\ref{ZZ_2e2mu_inv}). As
expected, there are two peaks in this distribution indicating the
first two excitations of the graviton considered in the RS model.
The transverse momentum (left) and rapidity (right) distributions
of the $e^+e^-$ pair and $\mu^+\mu^-$ pair are respectively shown
in Fig.\ (\ref{ZZ_2e_pt_rap}) and (\ref{ZZ_2mu_pt_rap}) with the 
insertion of $M^{e^+e^-\mu^+\mu^-}>600$ GeV cut. In these $P_T$
distributions the first kink is visible at the half value of the
first excitation. The rapidity distributions of those pairs are
not alike because of the same reason of applying aforesaid high
invariant mass cut for which the angular correlation between the
decay products of the two Z bosons has been lost. 

%%%%% WW plots begins %%%%%
\begin{figure}
 \centering{
 \includegraphics[height=8.5cm,width=7.4cm]{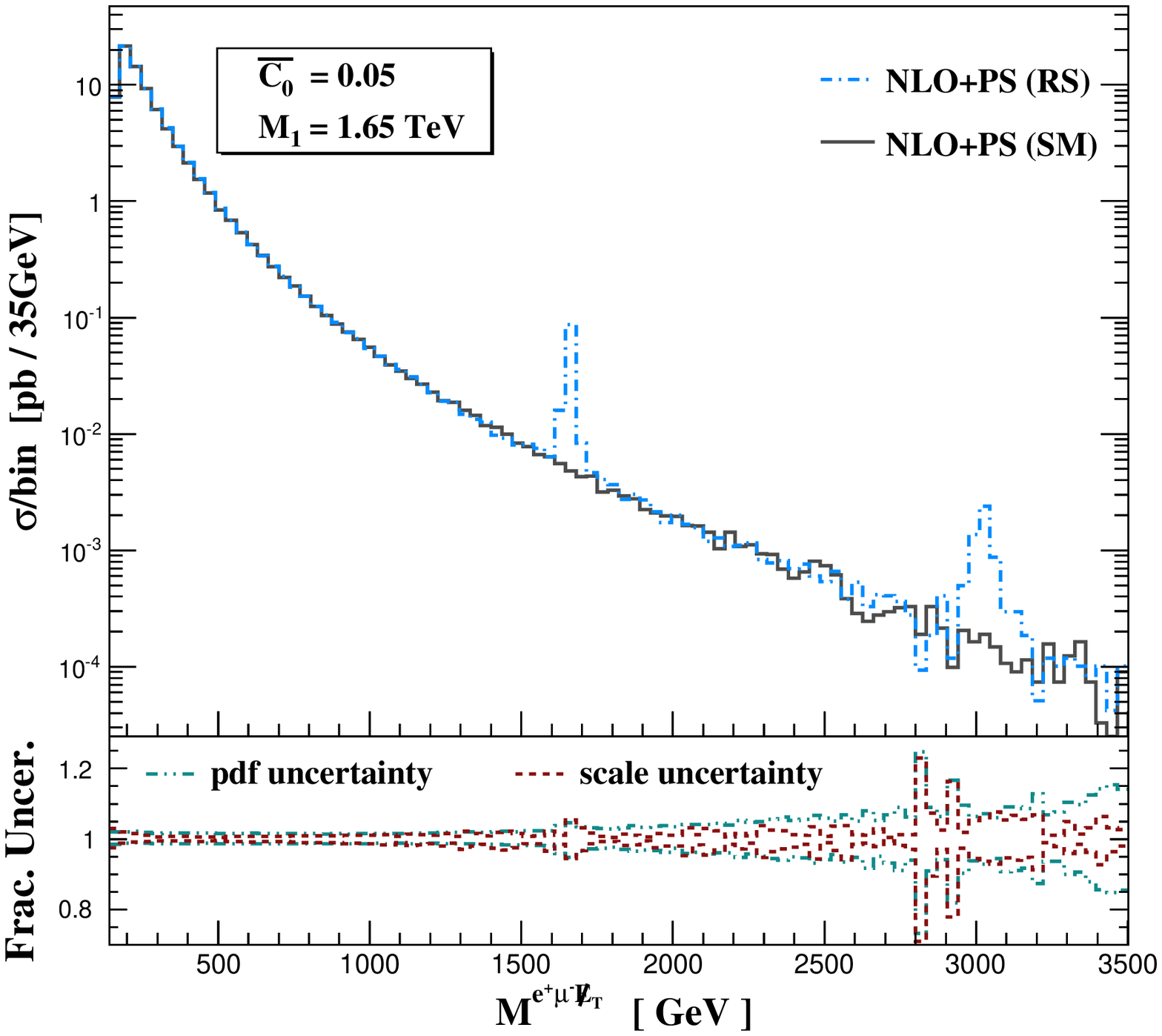}
 \includegraphics[height=8.5cm,width=7.4cm]{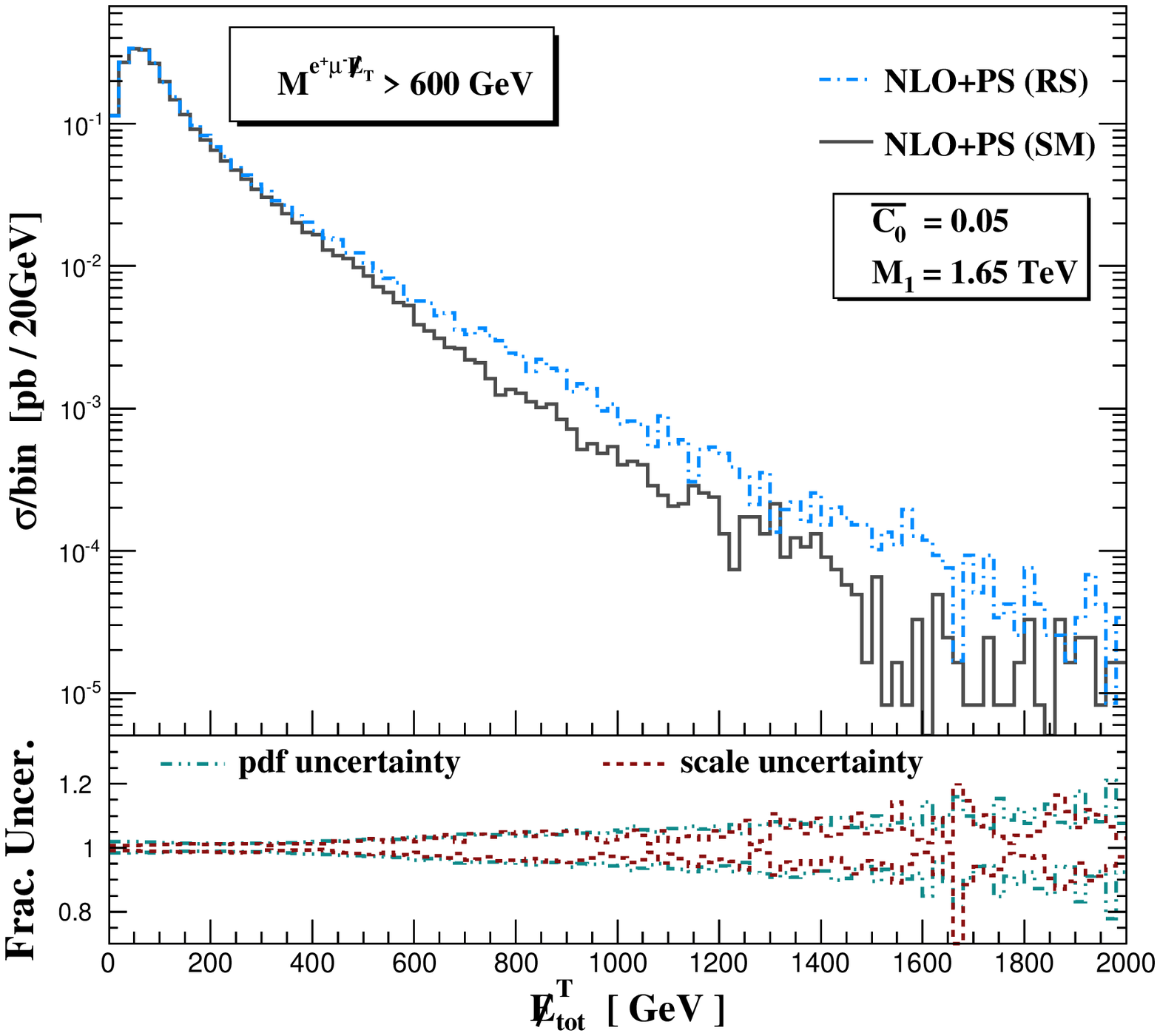}
 }
 \caption{Invariant mass (left) distribution of the decay products of $W^{\pm}$ and the total missing $\cancel{E}_T$ distribution (right) in SM and RS.}
 \label{WW_emuet_et_inv_et}
\end{figure}

\begin{figure}
 \centering{
 \includegraphics[height=8.5cm,width=7.4cm]{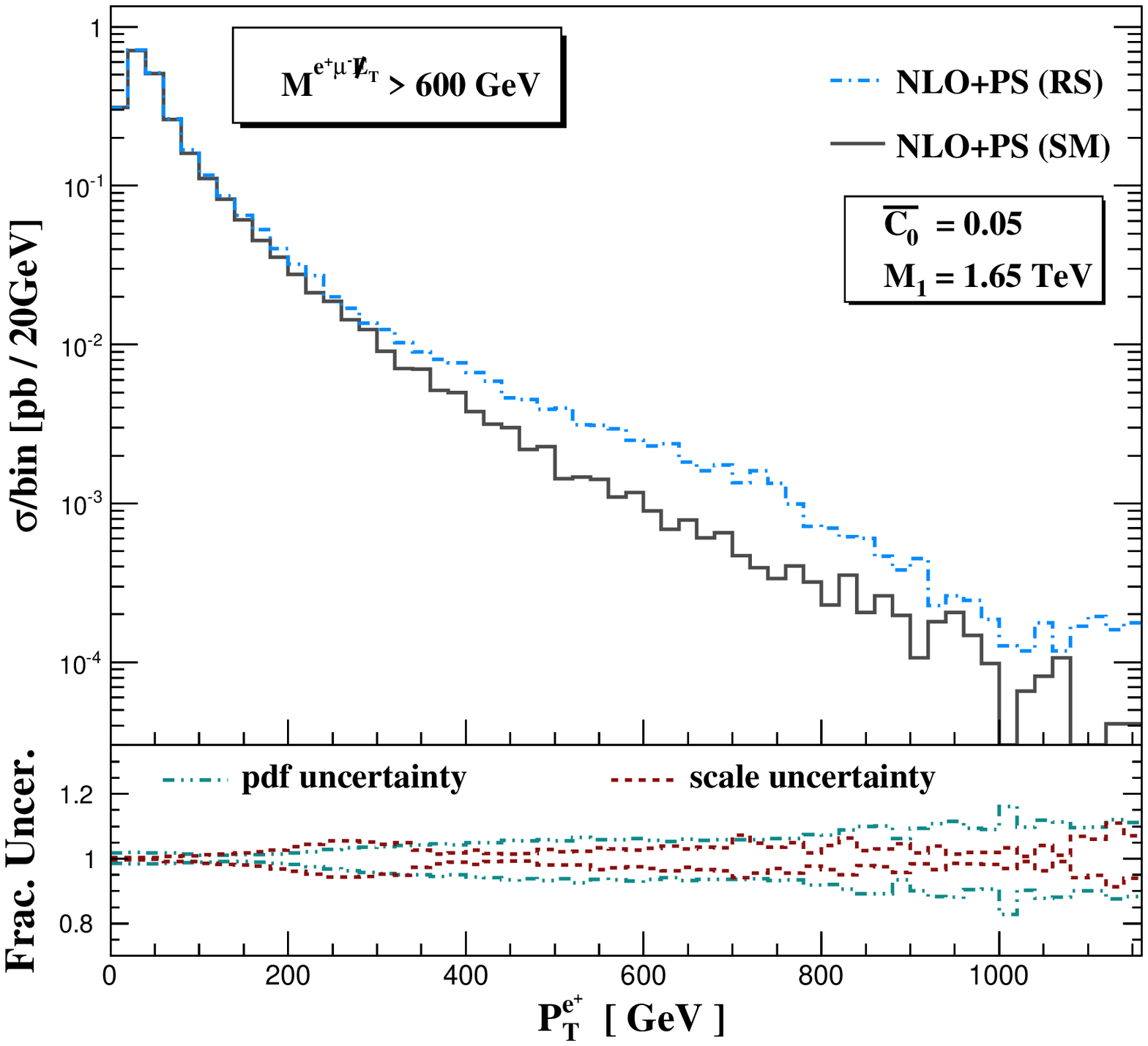}
 \includegraphics[height=8.5cm,width=7.4cm]{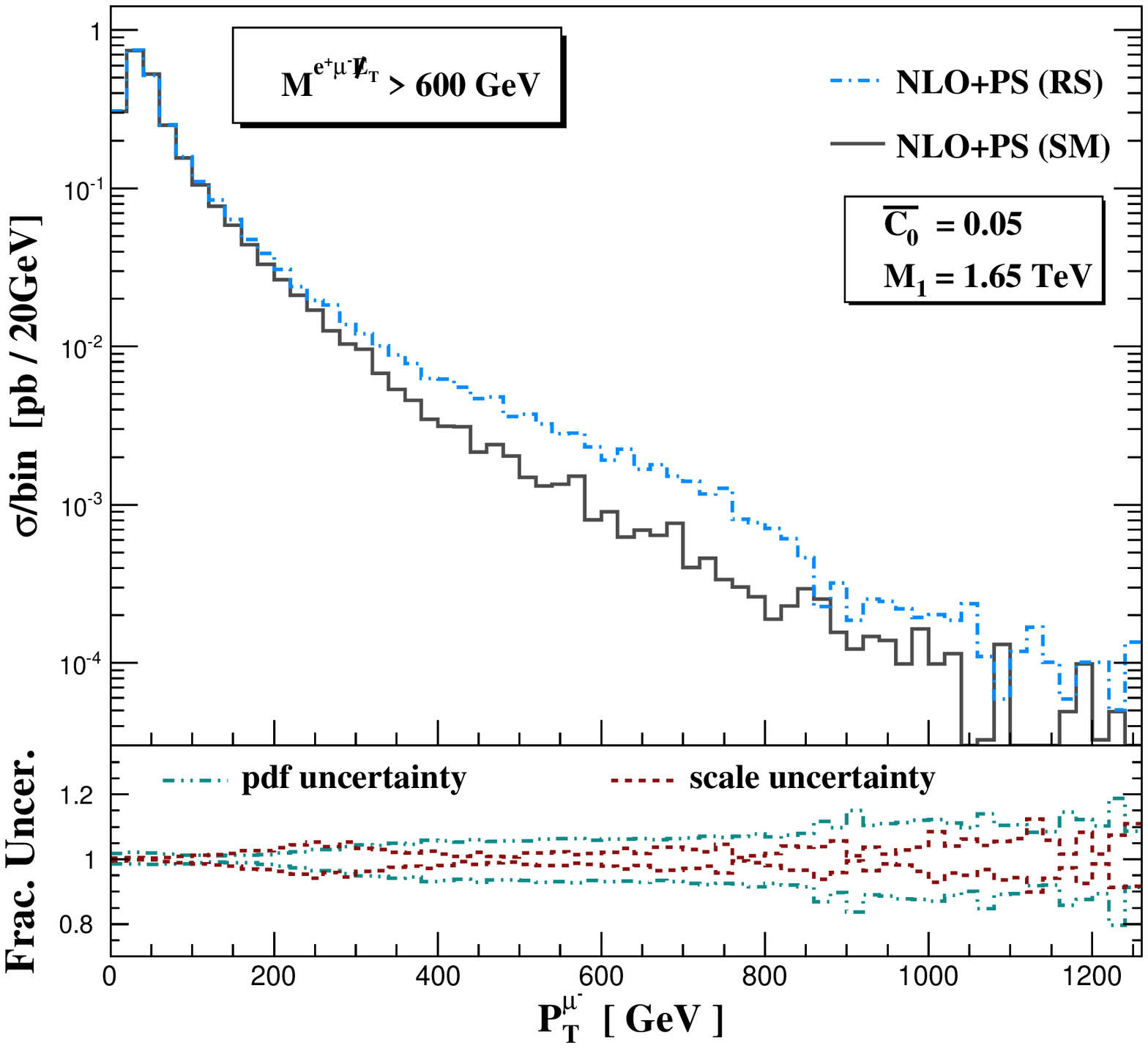}
 }
 \caption{Transverse momentum distribution of $e^+$ (left) and $\mu^-$ (right) coming from $W^{\pm}$ decay in RS and SM.}
 \label{WW_ep_mum_pt_pt}
\end{figure}
%%%%% WW plots ends %%%%%

Few selective distributions that are coming from $W^+W^-$ events are
given in Fig.\ (\ref{WW_emuet_et_inv_et}) and (\ref{WW_ep_mum_pt_pt}).
The left panel of Fig.\ (\ref{WW_emuet_et_inv_et}) shows the invariant
mass ($M^{e^+\mu^-\cancel{E}_T}$) distribution of the $W^{\pm}$ decay
products and in the right panel of this figure, we have presented the
transverse missing energy ($\cancel{E}_T$) distribution that is
coming from the electron neutrino and muon anti-neutrino which
practically escape the experimental detection in the collider.
$M^{e^+\mu^-\cancel{E}_T}>600$ GeV cut is used for the later one to
differentiate the missing $\cancel{E}_T$ signal of the RS from the SM
one. In Fig.\ (\ref{WW_ep_mum_pt_pt}), the transverse momentum
distributions of the positron (left) and the muon (right) are shown
in the region where $M^{e^+\mu^-\cancel{E}_T}>600$ GeV and we find
that the distribution in the RS case is comparatively harder than
the SM distribution in this region.  
\begin{table}[tbh]
\begin{center}
 \begin{tabular}{|c|c|c|c|c|}
  \hline
$\bar{c}_0$ &0.03 &0.05 &0.07 &0.10 \\ \hline
$M_1^{(3{\sigma})}$ (TeV) &4.5 &6.3 &7.4 &10.3 \\ \hline
$M_1^{(5{\sigma})}$ (TeV) &4.2 &5.2 &6.0 &8.3 \\ \hline
\end{tabular}
\end{center}
\caption{
Bounds on $M_1$ for various $\bar{c}_0$ values at the $14$ TeV LHC with
integrated luminosity of $50$ fb$^{-1}$
at 3-sigma and 5-sigma signal significance, coming from Drell-Yan process.
}
\label{lum_table_DY}
\end{table}

Here, we investigate the search sensitivity of the RS model using
the Drell-Yan and di-photon processes for the following $\overline{c}_0$
values at the 14 TeV LHC: $\overline{c}_0=0.03, 0.05, 0.07, 0.10$. We
have calculated the total cross section for the signal plus
background and the background alone using the invariant mass
distribution of the $e^+e^-$ ($\gamma\gamma$) pair in Drell-Yan
(di-photon) production and estimated the minimum required luminosity
that distinguishes the signal from the background at 3-sigma ($3\sigma$)
and 5-sigma ($5\sigma$) signal significance for various values of
$M_1$ for a fixed $\overline{c}_0$.  The required minimum luminosity is
defined as, $L_{\mbox{\scriptsize{min}}} = \mbox{max}\{L_{3\sigma(5\sigma)}, 
L_{3N_S(5N_S)}\}$, where $L_{3\sigma(5\sigma)}$ describes the
integrated luminosity at 3-sigma (5-sigma) signal significance and 
$L_{3N_S(5N_S)}$ denotes the integrated luminosity to have at least
$3(5)$ signal events.  From the data set of $M_1$ {\it vs.}\ 
$L_{\mbox{\scriptsize{min}}}$ thus prepared, by inversion we
find the $M_1$ value that corresponds to $50$ fb$^{-1}$ luminosity
for each of the $\overline{c}_0$ values listed above and those bounds
that are counted using Drell-Yan and di-photon processes are tabulated
in Table \ref{lum_table_DY} and \ref{lum_table_DP} respectively.  Of
course, a full analysis including the effects of detector simulation,
non-reducible backgrounds etc. would lead these bounds to their betterment.
\begin{table}[tbh]
\begin{center}
 \begin{tabular}{|c|c|c|c|c|}
  \hline
$\bar{c}_0$ &0.03 &0.05 &0.07 &0.10 \\ \hline
$M_1^{(3{\sigma})}$ (TeV) &5.2 &5.6 &6.1 &7.5 \\ \hline
$M_1^{(5{\sigma})}$ (TeV) &5.0 &5.3 &5.6 &6.4 \\ \hline
\end{tabular}
\end{center}
\caption{
Bounds on $M_1$ for various $\bar{c}_0$ values at the $14$ TeV LHC with
integrated luminosity of $50$ fb$^{-1}$
at 3-sigma and 5-sigma signal significance, coming from di-photon
production process.
}
\label{lum_table_DP}
\end{table}

\section{Conclusions}

In this paper, we have studied all the important di-final state
processes ($\ell^+ \ell^-$, $\gamma \gamma$, $ZZ$ and $W^+ W^-$) in
the RS model to NLO+PS accuracy, implemented in the {\sc a}MC@NLO
framework.  All the subprocesses to NLO in QCD have been taken into
account for both the SM and RS model.  For the di-final state processes
under consideration, we demonstrate the importance of NLO+PS results
over the fixed order NLO computations, by studying the $p_T$
distribution of the di-final states.  For suitable choice of RS model
parameters, a selection of the results for various observables at the
14 TeV LHC are presented.  PDF and scale uncertainties are presented for
the various distributions which significantly reduce with the inclusion
of NLO corrections.  The di-lepton and di-photon processes are used to
study the search sensitivity of the RS model at 14 TeV LHC with 50
fb$^{-1}$ luminosity.  The stand-alone codes can be used to generate
events with any choice of RS model parameters for di-final state
processes discussed in this paper to NLO+PS accuracy and are being
made available on the website \url{http://amcatnlo.cern.ch}.

\section*{Acknowledgements}
We would like to acknowledge the High Performance
Computing facility of Theory Division, SINP where the computational
work has been carried out. GD would like to thank Paolo Torrielli,
Marco Zaro and M. C. Kumar for important suggestions.

% \cleardoublepage
\bibliography{di-final-RS_v3}{}
\end{document}